\documentclass[aps,prl,twocolumn]{revtex4-1}
\usepackage{bm}
\usepackage{graphicx}
\usepackage{amsmath}
\usepackage{amssymb}
\usepackage[utf8]{inputenc}
\usepackage[T1]{fontenc}
\usepackage{color}
\usepackage{upgreek}
\usepackage{subfigure}
\usepackage[normalem]{ulem}
\usepackage[unicode=true,colorlinks=true,citecolor=blue]{hyperref}
\usepackage{placeins}
\usepackage{lipsum}
\usepackage{epsfig}
\usepackage[utf8]{inputenc}
\usepackage{makecell}
\usepackage{wrapfig}
\newcommand{\nix}[1]{}
\newcommand{\bl}[1]{\textcolor{blue}{#1}}
\newcommand{\re}[1]{\textcolor{red}{#1}}

\definecolor{greenI}{rgb}{0, .4, 0}

\usepackage[version=4]{mhchem}
\usepackage{glossaries}
\setacronymstyle{long-short}
\glsdisablehyper
\usepackage{siunitx}

\newcommand{\MilliMeter}[1]{\SI{#1}{\milli\metre}}
\newcommand{\MilliElectronVolt}[1]{\SI{#1}{\milli\electronvolt}}

\newcommand{\NanoMeter}[1]{\SI{#1}{\nano\meter}}
\newcommand{\NanoAmpere}[1]{\SI{#1}{\nano\ampere}}

\newcommand{\MilliWatt}[1]{\SI{#1}{\milli\watt}}

\newcommand{\TeraHertz}[1]{\SI{#1}{\tera\hertz}}

\newacronym{lpge}{LPGE}{linear photogalvanic effect}
\newacronym{cpge}{CPGE}{circular photogalvanic effect}
\newacronym{cpde}{CPDE}{circular photon drag effect}
\newacronym{lpde}{LPDE}{linear photon drag effect}
\newacronym{tblg}{tBLG}{twisted bilayer graphene}
\newacronym{hbn}{hBN}{hexagonal boron nitride}
\newacronym{pge}{PGE}{photogalvanic effect}
\newacronym{pde}{PDE}{photon drag effect}
\newacronym{cnp}{CNP}{charge neutrality point}
\newacronym{thz}{THz}{terahertz}
\newacronym{cw}{cw}{continuos wave}
\newacronym{qcl}{QCL}{quantum cascade laser}

\begin{document}

\title{Infrared photoresistance as a sensitive probe %\sout{of temperature variations}
of electronic transport in twisted bilayer graphene}

\author{S. Hubmann$^1$, G. Di Battista$^{2,3}$, I. A. Dmitriev$^1$, K. Watanabe$^4$, T. Taniguchi$^5$, D.K. Efetov$^{2,3,6}$, and S.D. Ganichev$^{1,7}$}

\affiliation{$^1$Terahertz Center, University of Regensburg, 93040 Regensburg, Germany}

\affiliation{$^2$ICFO - Institut de Ciencies Fotoniques, The Barcelona Institute of Science and Technology, Castelldefels, Barcelona 08860, Spain}

\affiliation{$^3$Fakultät für Physik, Ludwig-Maximilians-Universität, Schellingstrasse 4, 80799 München, Germany}

\affiliation{$^4$Research Center for Functional Materials, National Institute for Materials Science, 1-1 Namiki, Tsukuba 305-0044, Japan}

\affiliation{$^5$International Center for Materials Nanoarchitectonics, National Institute for Materials Science, 1-1 Namiki, Tsukuba 305-0044, Japan}

\affiliation{$^6$Munich Center for Quantum Science and Technology (MCQST), München, Germany}

\affiliation{$^7$CENTERA Laboratories, Institute of High Pressure Physics, Polish Academy of Sciences PL-01-142 Warsaw, Poland}

%\email{sergey.ganichev@ur.de}

\begin{abstract}
We report on observation of the infrared photoresistance of \gls{tblg} under continuous quantum cascade laser illumination at a frequency of \TeraHertz{57.1}. The photoresistance shows an intricate sign-alternating behavior under variations of temperature and back gate voltage, and exhibits giant resonance-like enhancements at certain gate voltages. The structure of the photoresponse correlates with weaker features in the dark dc resistance reflecting the complex band structure of \gls{tblg}. It is shown that the observed photoresistance is well captured by a bolometric model describing the electron and hole gas heating, which implies an ultrafast thermalization of the photoexcited electron-hole pairs in the whole range of studied temperatures and back gate voltages. We establish that photoresistance can serve a highly sensitive probe of the temperature variations of electronic transport in \gls{tblg}. 
\end{abstract}

%\pacs{
%}
\maketitle

%\textcolor{cyan}{Giorgio: My comments are in cyan}

\section{Introduction}
\label{introduction}

%\sdg{photoresistance}\re{Photoconductivity, -conductance -resistivity UNIFY!}

In a breakthrough discovery in 2018 it was experimentally shown that when two graphene layers are stacked vertically, while being twisted by a magic angle of 1.1$^\circ$, the inter-layer hybridization leads to the emergence of ultra-flat electronic bands~\cite{Bistritzer2011,Cao2018,Cao2018a}. Strikingly, these bands were found to host a plethora of exotic electronic phases including unconventional superconductivity, correlated insulators, as well as magnetic and topological phases ~\cite{Cao2018,Cao2018a,Yankowitz2019,Sharpe2019,Serlin2020}. These discoveries, demonstrating that the twist angle can be used to control the state of 2D materials and for manipulation of strong electronic correlations, produced a remarkable excitement and multi-disciplinary research (see, e.g., Refs.~\onlinecite{Andrei2020,Toermae2022,Song2022,Chaudhary2022,Jaoui2022,Grover2022,Cao2021,Hao2021,Wu2021,Andrei2021,Hesp2021a,Liu2020,uri2020,Cao2020,Balents2020,Otteneder2020,Yoo2019,Choi2019,Lu2019} and references therein).   
%\cite{Andrei2021,Balents2020,Otteneder2020,Jaoui2022,Grover2022,Hesp}
%Wu2018,Po2018,Xu2018a
%\cite{Wu2018,Po2018,Xu2018a,Yankowitz2019,Lu2019,Shen2020,Liu2020,Li2009,Yan2012a,Mishchenko2014,Santos2007,Morell2010,Mele2010,Bistritzer2011,Luican2011,Rozhkov2016,Kim2017a,Koshino2018,RibeiroPalau2018,Yoo2019,Kerelsky2019,Jiang2019,Serlin2020,Xie2020,Stepanov2020,Cao2020,Lu2021,Bernevig2021,Cao2021,Hao2021,Das2021,Wu2021,Morell2010,Choi2019,Otteneder2020,Lu2021,Wu2021,Jaoui2022,Grover2022} 
%\re{do not cancel our nano letters and add some new papers 2022,2021}. 
In particular, optoelectronic studies provide a way to access the unique and rich physics of \gls{tblg} and open a potential for a novel kind of devices, such as detectors of terahertz and infrared radiation \cite{Seifert2020, DiBattista2021}. So far, these studies have been limited to investigation of photocurrents excited in unbiased \gls{tblg} structures \cite{Xin2016,Yin2016a,Otteneder2020,Sunku2020,Hesp2021,Sunku2021,Hubmann2022,Deng2020,KortKamp2018,Gao2020a,Wang2020,Chen2022,Zheng2022}. 

Here we report on observation and study of the infrared photoresistance in \gls{tblg} with small twist angle of $ \sim 1^\circ$. The  photoresistance exhibits a complex sign-alternating behavior upon variation of the back gate voltage and the sample's temperature.
In particular, at low temperatures it exhibits sharp negative spikes at several gate voltages.
%Furthermore, it is enhanced drastically and resonance-like at several gate voltages. 
We show that the infrared photoresistance is caused by the bolometric effect - change of the dc resistivity due to electron and hole gas heating. This conclusion is supported by the fact that the photoresistance closely follows the first derivative of the dark dc resistance with respect to the temperature: the shape of the gate-voltage dependence of the photoresistivity at all temperatures practically coincides with the difference of resistance traces obtained at %slightly different 
two neighboring temperatures. 
%temperature, such that its gate-voltage dependence practically coincides with the difference of resistance traces obtained at slightly different temperatures. 
This property enables us to estimate the heating effect and to demonstrate its weak sensitivity both to the measurement temperature and to the gate voltage, controlling the low-energy moiré band structure and position of the chemical potential in \gls{tblg}. We thus establish that, despite high energy of electron-hole pairs excited by the infrared radiation, the rich structure and sharp negative spikes in the gate voltage of observed photoresistance are completely determined by the temperature variations of low-energy transport, which is strongly affected by the moiré potential of small angle \gls{tblg}. Our study demonstrates that photoresistance can serve as a sensitive probe of the low-energy transport characteristics, even those that are hardly detectable using the standard transport measurements.
%The position and shape of the observed negative spikes of the photoresistance 
%resonances \sh{(can we call this resonance? I would rather speak about peaks)} 
%thus reflect weaker variations of the sample resistance caused by the moiré band structure of small angle \gls{tblg}. 
%Our measurements demonstrate that photoresistance can be a very sensitive probe of the temperature variations of the transport characteristics, even those that are hardly detectable using the standard transport measurements.

\section{Samples and methods}
\label{samples_methods}

The heterostructure consisting of the \gls{tblg} layer sandwiched between \gls{hbn} was prepared using a ''cut and stack'' technique, see Ref.~\cite{Kim2017a}. Bottom and top layers of \gls{hbn} had a thickness of 16 and \NanoMeter{10}, respectively. The two graphene sheets were stacked one on top to the other at a  target angle of $\SI1{\degree}$. 
%\textcolor{cyan}{\sout{and consequently relaxed into the \re{final angle $\theta\approx ???$ (Giorgio please update)}}}. 
A graphite layer on the bottom of the \gls{hbn}/\gls{tblg}/\gls{hbn} was used as a local back gate electrode.  The stack was etched into a Hall bar geometry 17$\times$ \SI{2}{\micro\meter\squared}. A \ce{CHF_3}/\ce{O_2} mixture was used to expose the graphene edges, with subsequent  evaporation of \ce{Cr}/\ce{Au} (5/\NanoMeter{50}) providing ohmic contacts for transport and photoresistance measurements.

The sample was placed in a temperature-variable \ce{He} exchange gas optical cryostat with \ce{ZnSe} windows. To capacitively tune the carrier density in the \gls{tblg} structure, back gate voltage $U_{\text{G}}$ in the range of $\pm \SI3{\volt}$ was applied to the graphite back gate of the device. The sample resistance, $R$, was measured in two-terminal geometry using the standard low-frequency lock-in technique, with excitation current of \NanoAmpere{100}. 

\begin{figure}
	\centering
	\includegraphics[width=\linewidth]{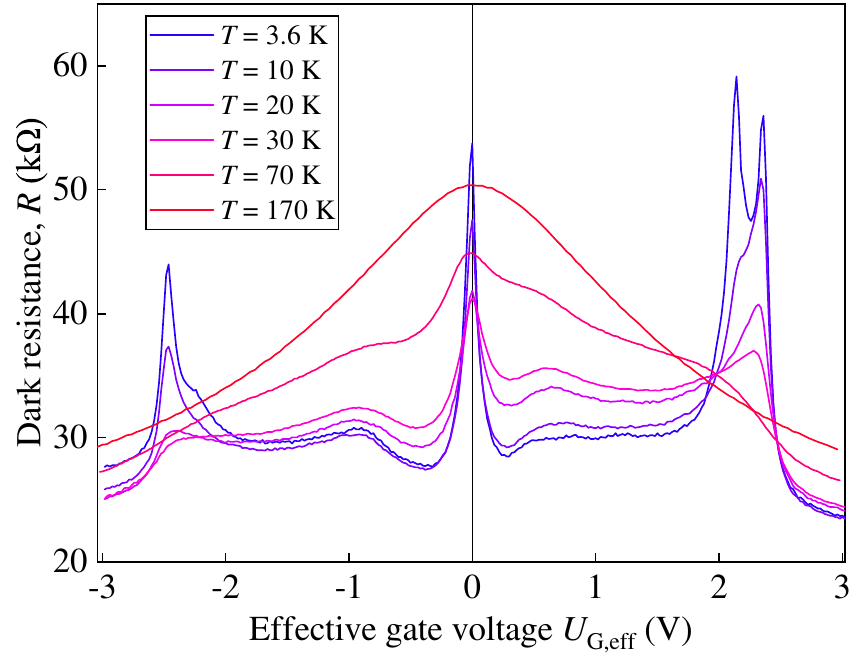}
	\caption{Two-point dc resistance $R$ measured between contacts B and C (see inset in  Fig.~\ref{Fig2}) as a function of the effective gate voltage at temperatures $T$ ranging from 3.6 to \SI{170}{\kelvin}.}
	\label{Fig1}
\end{figure}

Fig.~\ref{Fig1} shows the sample resistance as a function of the applied gate voltage measured at different temperatures in the range from $T=$3.5 to \SI{170}{\kelvin}. At low temperatures the resistance exhibits clear sharp peaks at certain gate voltages characteristic for \gls{tblg}: The \gls{cnp}-peak at $U_\text{G,eff}=0$ V is flanked by highly-resistive peaks at $\pm \SI{2.5}{\volt}$ with strongly insulating behavior. From the position of these peaks which mark the edges of the moiré bands, the twist angle was estimated to be $\sim \SI1{\degree}$, using the thickness of the bottom hBN layer to calculate the gate capacitance. The resistance traces also show less pronounced peaks at $\pm \SI1{\volt}$ suggesting that the area between the source and drain contacts B and C (see Fig.~\ref{Fig2}) acquires some twist angle inhomogeneity \cite{uri2020}, i.e., that the measured area is dominated by a twist angle of $\sim \SI1{\degree}$ but also contains fractions with lower twist angles of $\sim \SI{0.6}{\degree}$. The inhomogeneity can also be responsible for the double-peak structure of $R$ at $\pm \SI{2.5}{\volt}$. Note that the gate voltage corresponding to the \gls{cnp}, $U_{\text{CNP}}$, was varying slightly between different sample cooldowns due to different charge trapping in the gate insulator \cite{Dantscher2017,Candussio2021}. Correspondingly, in the presented data we use the effective gate voltage $U_\text{g,eff}=U_\text{G}-U_\text{CNP}$.

In order to measure the photoresistance we used a \gls{cw} \gls{qcl} which operated at a radiation frequency of \TeraHertz{57.1} (photon energy of \MilliElectronVolt{236})  and provided an maximum output power of \MilliWatt{130}. The normally incident radiation was focused onto the \gls{tblg} sample using a parabolic mirror, which resulted in a laser spot with diameter of about \MilliMeter{0.5} as checked by a pyroelectric camera \cite{Ziemann2000,Drexler2012}. This spot diameter was an order of magnitude larger than the Hall bar size ensuring a uniform illumination of the sample. The polarization state of the incoming radiation was controlled using a quarter-wave plate and linear polarizers. The photoresistance was measured as the difference between the two-point dc resistance in the presence and absence of \gls{cw} \gls{qcl} illumination, $\Delta R=R_{\text{ill}}-R$. In all figures apart from Figs.~\ref{Fig5} and \ref{Fig6}, we present the results for the normalized photoresistance $\Delta R/R$.

\section{Results}
\label{results}

\begin{figure}
	\centering
	\includegraphics[width=\linewidth]{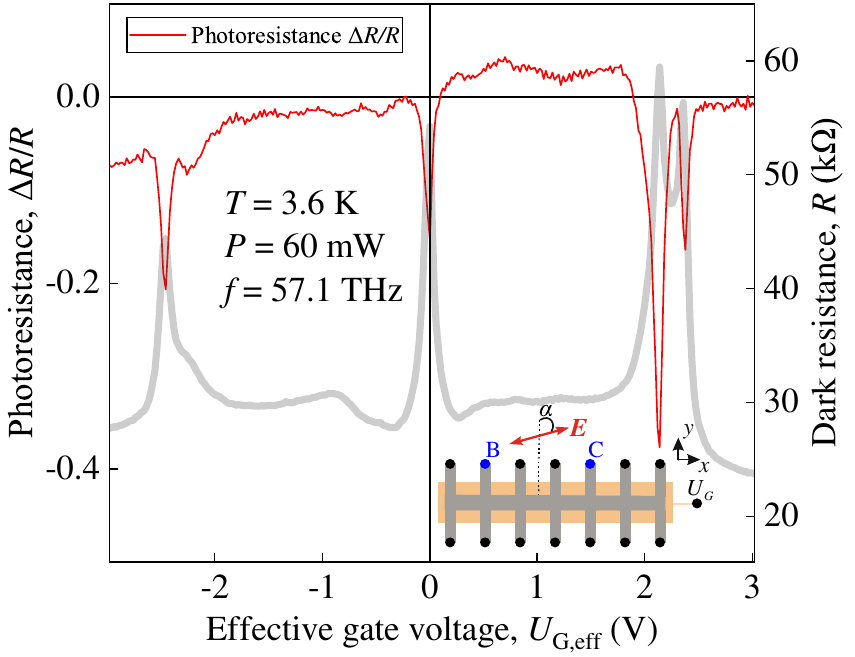}
	\caption{Normalized photoresistance $\Delta R/R$ as a function of the effective gate voltage at a temperature $T =$ $\SI{3.6}{\kelvin}$. The  corresponding resistance without illumination is illustrated by thick gray line. The inset shows a sketch of the Hall bar structure, contacts B and C used for the two-point measurements, and the linear polarization angle with respect to the short side of the Hall bar (the azimuth angle $\alpha=75^\circ$).
	% between the short side of the Hall bar and polarization direction of the radiation electric field $\bm E$ equal to 75$^{\circ}$).
	}
	\label{Fig2}
\end{figure}

Applying the infrared radiation to the \gls{tblg} structure we observed a photoinduced change $\Delta R$ of the sample resistance \footnote{Similar to our previous works \cite{Otteneder2020,Hubmann2022} (implementing much lower frequencies), in the absence of external bias we also observed polarization-dependent photocurrents. The corresponding results are presented in 
%Supplementary Material, see 
Sec.~\ref{appendix}}.
%Note that the illumination of the \gls{tblg} structure also results in the formation of a polarization-dependent photocurrent. The photocurrent was discussed in previous works \cite{Otteneder2020,Hubmann2022} and, therefore, the photocurrent results are presented in the appendix, see Sec.~\ref{appendix}.}. 
Fig.~\ref{Fig2} shows a typical example of recorded $\Delta R$, normalized to the dark resistance $R$, as a function of the applied gate voltage $U_\text{G,eff}$. These data were obtained at the lowest $T=$\SI{3.6}{\kelvin}. Besides a primary narrow negative spike at the \gls{cnp}, the photoresistance $\Delta R/R$ exhibits several spikes at large negative and positive gate voltage, namely, at  $U_{\text{g,eff}}=-2.46$, $U_\text{g,eff}=2.14$, and $\SI{2.38}{\volt}$. Comparison of these data with the dark resistance, see gray line in Fig.~\ref{Fig2}, shows that positions of the negative spikes in $\Delta R/R$ coincide with the peak positions in $R$. We thus observe that, despite the huge photoexcitation energy of \MilliElectronVolt{236}, the major sharp features in the $U_\text{g,eff}$-dependence of $\Delta R/R$ correspond to the abrupt changes of dc transport properties when the chemical potential is passing the edges of the low-energy moiré bands at $\sim$\MilliElectronVolt{10} from the \gls{cnp}, see discussion below. This establishes that the infrared photoresistance can indeed serve a sensitive probe of the low-energy dc transport of \gls{tblg}. In this connection, it is worth mentioning that variations of the photoresistance $\Delta R$ with $U_\text{g,eff}$ are orders of magnitude stronger than those in the dark resistance $R$ (not exceeding 50 \%), and thus remain equally strong if the photoresistance $\Delta R$ is not normalized to the dark resistance $R$, see Fig.~\ref{Fig5}.

\begin{figure}
	\centering
	\includegraphics[width=\linewidth]{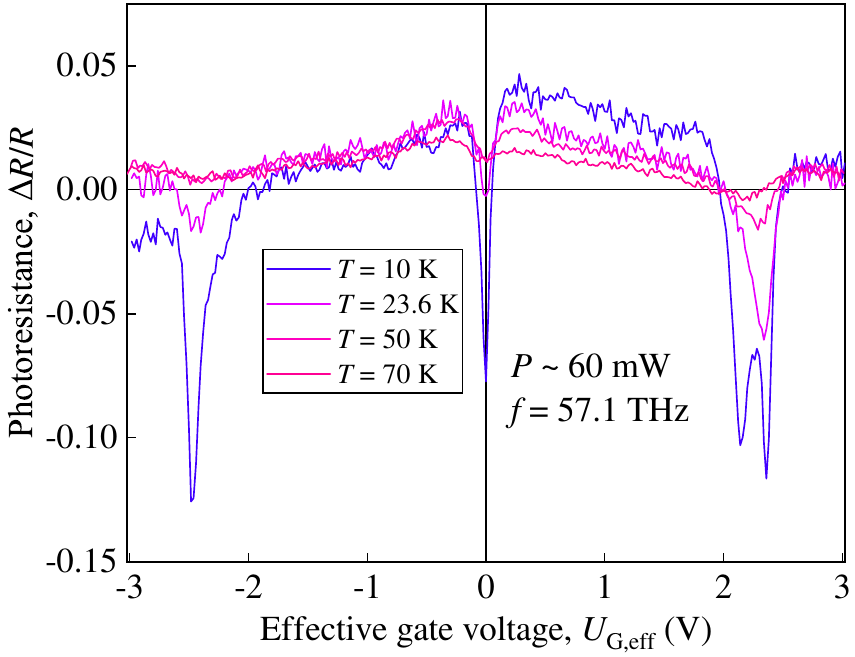}
	\caption{Normalized photoresistance $\Delta R/R$ as a function of the effective gate voltage at different temperatures from 10 to \SI{70}{\kelvin}.}
	\label{Fig3}
\end{figure}

\begin{figure}
	\centering
	\includegraphics[width=\linewidth]{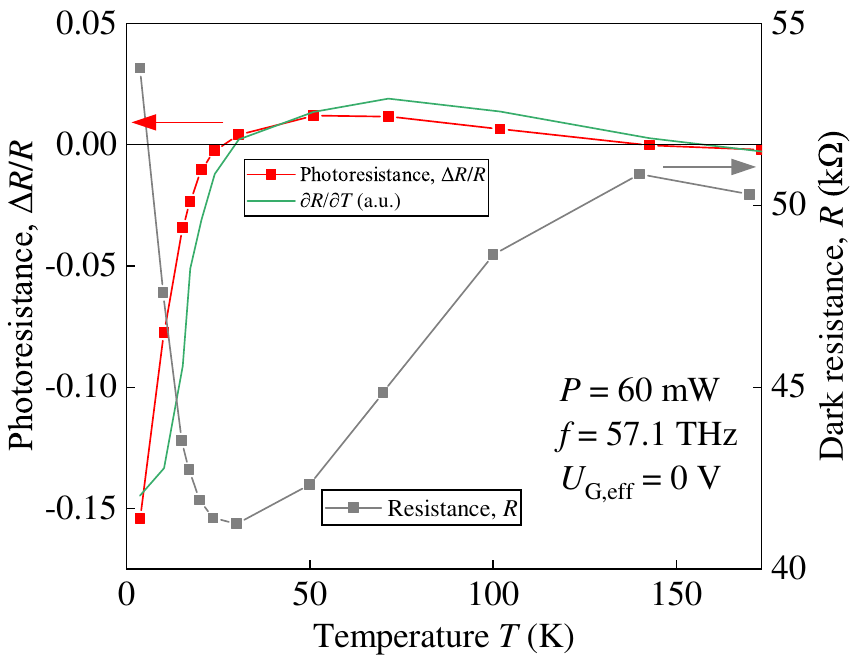}
	\caption{Temperature evolution of normalized photoresistance $\Delta R/R$ (red), resistance $R$ (gray), and the derivative of the resistance with respect to the temperature $\partial R/$$\partial T$ (green) at an effective gate voltage of $U_{\text{g,eff}}=\SI0{\volt}$ corresponding to the \gls{cnp}.}
	\label{Fig4}
\end{figure}

\begin{figure}
	\centering
	\includegraphics[width=\linewidth]{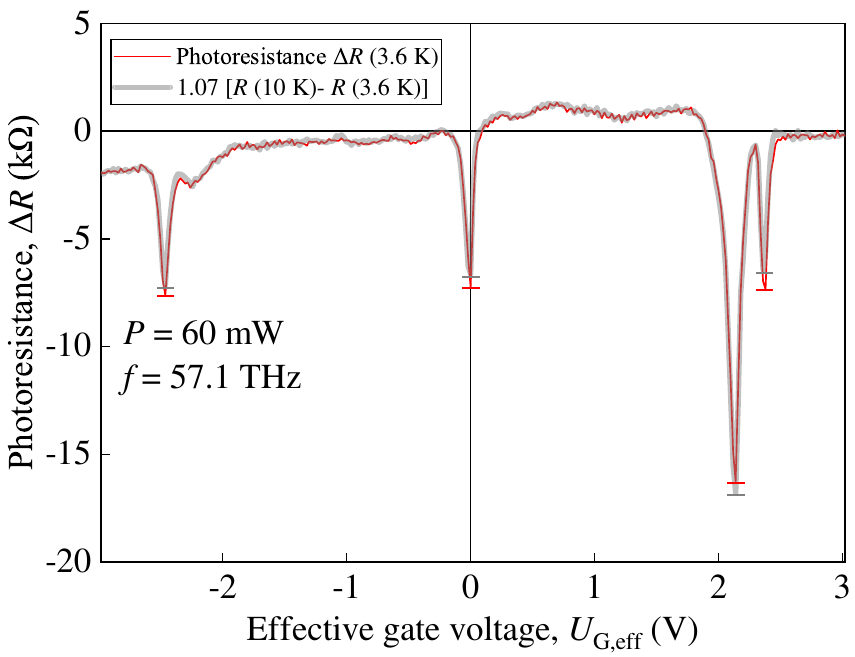}
	\caption{Photoresistance $\Delta R$ as a function of the effective gate voltage measured at a temperature of $T = \SI{3.6}{\kelvin}$ plotted together with the scaled difference between the dark resistances measured at $T = \SI{10}{\kelvin}$ and $T=\SI{3.6}{\kelvin}$.}
	\label{Fig5}
\end{figure}

\begin{figure*}
	\centering
	\includegraphics[width=\linewidth]{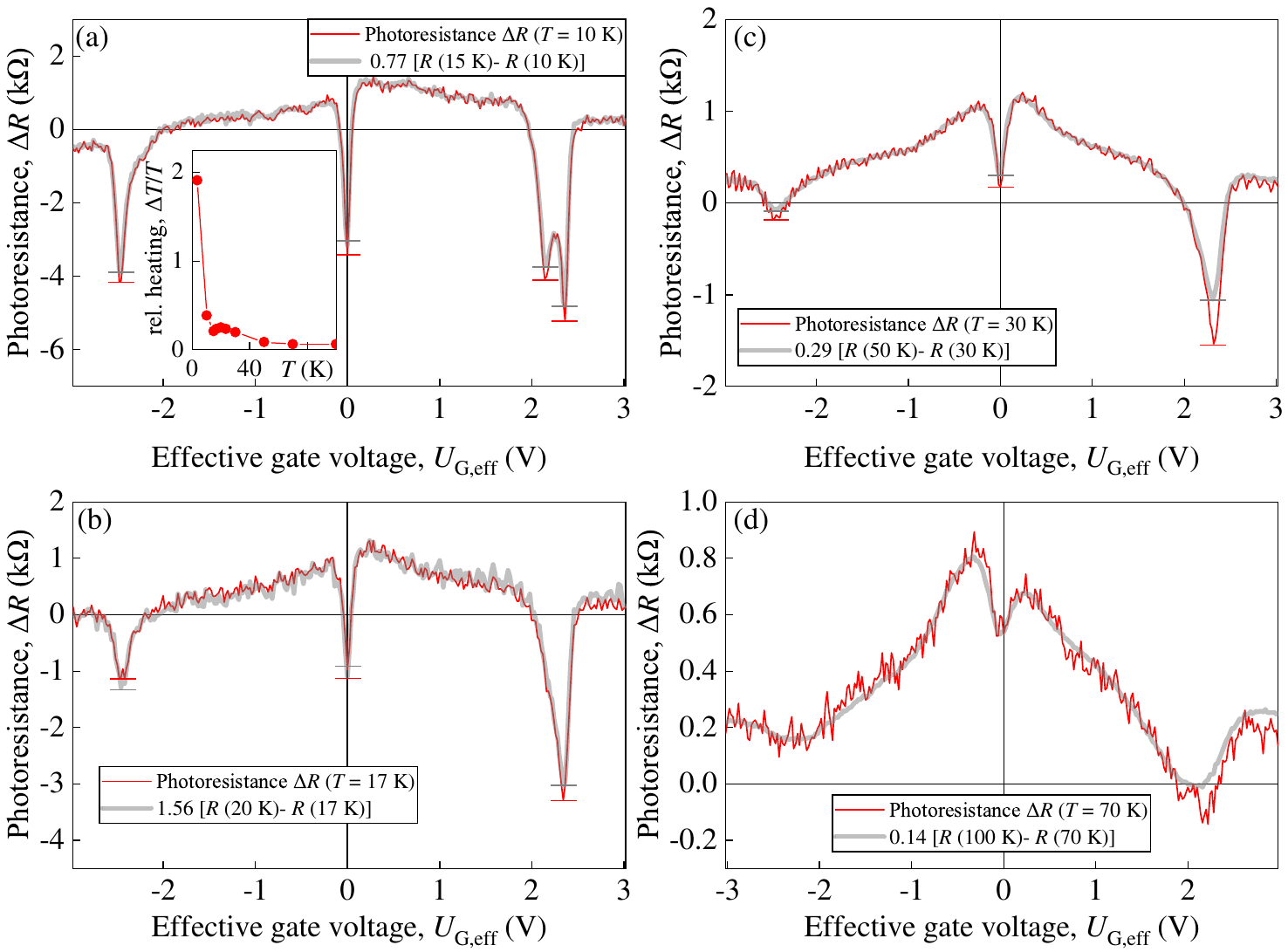}
	\caption{Photoresistance $\Delta R$ as a function of the effective gate voltage measured at temperatures of \SI{10}{\kelvin} (a), \SI{17}{\kelvin} (b), \SI{30}{\kelvin} (c), and \SI{70}{\kelvin} (d) plotted together with the scaled difference between the dark resistances measured at neighboring temperatures (provided in the legends together with the scaling factors).
	%$\SI{15}{\kelvin} $ and $T=\SI{10}{\kelvin}$ (a), $\SI{20}{\kelvin} $ and $T=\SI{17}{\kelvin}$ (b), $\SI{50}{\kelvin} $ and $T=\SI{30}{\kelvin}$ (c), and $\SI{100}{\kelvin} $ and $T=\SI{70}{\kelvin}$ (d). The corresponding scaling factors are given in the legends.
	The inset shows the relative radiation-induced heating $\Delta T/T$ estimated from the scaling factors using Eq.~\eqref{eq1}. 
	%plotted against the temperature.
	}
	\label{Fig6}
\end{figure*}

As the temperature of measurements increases, the behavior of the photoresistance changes substantially, see Fig.~\ref{Fig3}. While the photoresistance dips at positions of the peaks in dark resistance remain pronounced up to \SI{70}{\kelvin}, their magnitude is significantly reduced with increasing $T$. These changes are again consistent with evolution of the dc resistance $R$ in Fig.~\ref{Fig1}, where all peaks become progressively weaker and broader at elevated $T$ (for $T\lesssim\SI{30}{\kelvin}$). Moreover, one observes that the sign of $\Delta R$ at a fixed $U_\text{G,eff}$ follows the sign of temperature variation of $R$ at the same $U_\text{G,eff}$, for a detailed comparison see Figs.~\ref{Fig4}, \ref{Fig5}, and \ref{Fig6}. At higher $T$, the side peaks become very weak in $R$ while the corresponding features are still well resolved in $\Delta R$, whereas the broad \gls{cnp} peak in $R$ continues to broaden and starts to grow. Consistently, at $T>\SI{70}{\kelvin}$ the photoresistance becomes positive in the whole range of gate voltages (see Fig.~\ref{Fig6}(d). Finally, for temperatures above $\SI{140}{\kelvin}$ the photoresistance signal becomes vanishingly small.

We now turn to a quantitative comparison of the temperature evolution of the dc resistance and photoresistance. Figure \ref{Fig4} shows the temperature dependence of the normalised photoresistance $\Delta R/R$ (red line) together with that of the resistance $R$ without illumination (gray) at  $U_{\text{G,eff}}=0$ corresponding to the \gls{cnp}. It is clearly seen that the photoresistance changes its sign with increasing temperature. Apart from that, comparing the temperature dependence of the photoresistance with the dark resistance we find that the photoresistance closely follows the temperature derivative of the dark resistance, $\partial R/$$\partial T$, see green line in Fig.~\ref{Fig4}. As discussed below, this provides a strong evidence that the observed photoresistance is caused by the radiation-induced electron and hole gas heating. Another justification comes from the analysis in Fig.~\ref{Fig5}. Here we directly compare the non-normalized photoresistance $\Delta R$ measured at $T=$\SI{3.6}{\kelvin} (red line) with the difference of the dark resistance traces measured at temperatures of 10 and \SI{3.6}{\kelvin} and observe that, up to a constant scaling factor of 1.07, these curves nearly coincide, see Fig.~\ref{Fig5}. Similar precise coincidence is observed at higher $T$, see Fig.~\ref{Fig6}.

%1. Photoresistance is measured as a function of the gate voltage (positive, negative bias, delta R) [Footnote, references: photocurrent is observed as well (polarization dependent) Fig. 8-->10, previously discussed in 2 papers, similar explanation,  therefore only presented in Supplementary]
%
%2. multiple peaks, opposite signs in different regions
%
%3. Temperature changes peak magnitudes, results in the sign change
%
%4. T-dependence measured at one of the peaks --> change of sign; in agreement with resistance temperature dependent first derivity (plot it as well?) --> photoconductivity due to electron gas heating
%
%5. IR Delta R vs Delta R for two different temperatures (Now two versions, let's stay with the second one Fig.6 and move the first one to Supplem)
%
%6.  Fig. 7 no corresponding IR data, therefore, move to the Suppl.
%
%7. Very short discussion or even only a summary 

\section{Discussion}
\label{discussion}

The results presented above demonstrate that, despite a complex moiré band structure of \gls{tblg}, the photoresistance of this material in the studied frequency range is pretty well captured by a rather common and well-established mechanism related to electron heating \cite{Ganichev2005}. Within this mechanism, the stationary non-equilibrium energy distribution of electrons under continuous illumination is approximately given by the equilibrium Fermi-Dirac distribution, but with the measurement temperature $T$ replaced by an elevated electron temperature $T_\text{e}>T$. The value of the electron temperature $T_\text{e}$ should be found self-consistently from the energy balance equation. This equation expresses the stationary condition that, for certain $T_\text{e}>T$, the energy absorbed by electrons is fully compensated by the energy flow from hot electrons to the lattice (usually assumed to remain at the measurement temperature $T$). Provided $ T_e-T\ll T$, the photoresistance due to electron heating is given by
\begin{equation}\label{eq1}
\Delta R= \dfrac{\partial R}{\partial T_\text{e}}(T_\text{e}-T)\,,
\end{equation}
in full accordance with our findings, presented in Figs.~\ref{Fig4}-\ref{Fig6} and discussed in more details below. This description of electron heating and of the corresponding photoresistance is generally valid when equilibration of the absorbed energy within the electron system is faster than its transfer to the thermal bath of phonons but, in practice, is also frequently applicable when this condition is violated.

In our case, the photon energy, $\hbar\omega=\MilliElectronVolt{236}$, strongly exceeds all other involved energy scales - the temperature, Fermi energy, and moir\'e minibands widths -- all being of the order of 1 to \MilliElectronVolt{10} \cite{Cao2018}. It follows that a typical optical absorption process takes place between occupied initial electron states well ($\sim$ \MilliElectronVolt{120}) below the Fermi energy $E_\text{F}$ and empty states well above $E_\text{F}$. At such high energies the states are only weakly influenced by the moiré superlattice \cite{DiBattista2021} and can be considered as a continuum without significant modulations of the density of states. Thus, a change of the gate voltage, which strongly modifies the moiré electron spectrum and resistance in the vicinity of Fermi energy and \gls{cnp}, should only weakly affect the amount of absorbed energy governed by such distant states. Similarly, the relaxation of hot electrons should also possess a weak sensitivity to $U_\text{g}$. The photoexcited electrons and holes are expected to rapidly thermalize via electron-electron collisions simultaneously transferring the excess energy to the lattice via the acoustic phonon emission \footnote{Note that the energy of photoexcited electrons and holes ($\sim$ \MilliElectronVolt{120}) is still insufficient for emission of optical phonons having higher energy of $\sim$ \MilliElectronVolt{200}.}. All these processes involve huge number of possible intermediate states with typical energies comparable to $\hbar\omega\gg E_\text{F}$ and, thus, may also possess a weak sensitivity to the low-energy spectrum of \gls{tblg} and the exact position of $E_\text{F}$, controlled by the gate voltage $U_\text{g}$.

As a result, the electron temperature $T_\text{e}$, expressing the balance between absorption and energy relaxation, is expected to be largely insensitive to $U_\text{g}$. In sharp contrast to $T_\text{e}$, the temperature derivative ${\partial R}/{\partial T_\text{e}}$
of the dark resistance should be highly sensitive to both $U_\text{g}$ and temperature, in particular in the vicinity of the moiré band edges where the transport contributions of different states in the temperature window around the Fermi energy can be essentially different.  

The above conclusions are well supported by our experimental findings, which confirm that the sign of the photoresistance correlates with the temperature dependence of $R$, see Fig.~\ref{Fig4} \footnote{In general, the resistance $R$ depends on both electron and lattice temperatures, but at low $T$ the latter dependence is expected to be weak due to the lack of thermal phonons with relevant momenta.}. Moreover, the analysis presented in Fig.~\ref{Fig5} confirms that the $U_\text{g}$-dependence of $\Delta R$ coincides with that of ${\partial R}/{\partial T_\text{e}}$, while the electron heating factor $\Delta T\equiv T_e-T$ in Eq.~(\ref{eq1}) turns out to be insensitive to $U_\text{g}$. This implies a surprisingly low sensitivity of the high-frequency heating to the low-energy spectrum of \gls{tblg}. As discussed above, such a weak sensitivity is expected for the photoexcitation process, as well as the initial stages of thermalization and energy transfer to lattice. However, at a later stage, when the energy of nonequilibrium carriers reduces to $\sim$ \MilliElectronVolt{10}, one would rather expect that the relaxation process becomes sensitive to details of the band structure and position of the chemical potential. Nevertheless, our observations suggest that in the studied device the thermalization remains ultimately fast under all conditions, resulting in $T_e$ independent of $U_\text{g}$. 

This result is further confirmed by the analysis in Fig.~\ref{Fig6}: At all temperatures the shape of the photoresistance can be well reproduced by the difference of the dark resistance traces measured at two neighboring temperatures. Only in close vicinity of the strong negative spikes some small and not systematic deviations are noticeable (see horizontal bars in Figs.~\ref{Fig5} and \ref{Fig6}). These can be attributed by a limited accuracy of estimated $\partial R/\partial T$ obtained from comparison of resistance traces at two different temperatures. The scaling coefficients, obtained from such comparison (see legends in Figs.~\ref{Fig5} and \ref{Fig6}) provide an estimate for the electron heating $\Delta T$, which remains at the level of 5 K in the whole studied interval of temperatures. 
%The scaling coefficients \sh{\sout{, obtained from such fits}}  \sh{(These are determined by the gate voltage average/median of the quotient between Photoresistance and resistance difference. I would not call this a fit.)} (see legends in Figs.~\ref{Fig5} and \ref{Fig6}) provide an estimate for the electron heating $\Delta T$, which remains at the level of 5 K in the whole studied interval of temperatures. \sh{(should we mention here in some note that we consider $partial R/\partial T$ as constant between corresponding temperatures for this estimation)} 
From this estimate, we also establish that the relative heating $\Delta T/T$ (see inset in Fig.~\ref{Fig6}) remains small for all $T$ except the lowest $T=\SI{3.6}{\kelvin}$ (Fig.~\ref{Fig5}). In the latter case Eq.~(\ref{eq1}), valid for the linear heating regime, is only marginally applicable, and the obtained value of $\Delta T$ can be inaccurate.

\section{Summary}
\label{summary}

Summarizing, we show that, despite high energy of electron-hole pairs excited by the infrared radiation, the rich structure and sharp negative spikes in the gate voltage dependence of observed photoresistance are fully determined by the temperature variations of low-energy transport and, therefore, are strongly affected by the moiré potential of small angle \gls{tblg}. Our main observations and analysis demonstrate that photoresistance provides an alternative highly sensitive method for characterization of low-energy transport properties of \gls{tblg} which, despite its intrinsic complexity, permits a reliable treatment and clear understanding. In addition, the analysis of resistance detected in the presence and absence of radiation at varying $T$ yields a direct access and measure of electron heating in illuminated \gls{tblg}, which may provide an important quantitative check to future theories describing optical excitation and relaxation processes in this intriguing and rapidly developing class of 2D electronic systems.

\section{Acknowledgments}
\label{acknow} 
The support from the Deutsche Forschungsgemeinschaft (DFG, German Research Foundation) via Project SPP 2244 (GA501/17-1) and project DM1-5/1 (I.A.D.), and from the Volkswagen Stiftung Program (97738) are gratefully acknowledged. S.D.G. thanks the support from the IRAP program of the Foundation for Polish Science (grant MAB/2018/9, project CENTERA). D.K.E. acknowledges support from the Ministry of Economy and Competitiveness of Spain through the ‘Severo Ochoa’ programme for Centres of Excellence in R and D (SE5-0522), Fundacio Privada Cellex, Fundacio Privada Mir-Puig, the Generalitat de Catalunya throughthe CERCA programme, funding from the European Research Council (ERC) under the European Union’s Horizon 2020 research and innovation programme (grant agreement no. 852927). G.D.B. acknowledges support from the “Presidencia de la Agencia Estatal de Investigación” (Ref. PRE2019-088487). K.W. and T.T. acknowledge support from JSPS KAKENHI (Grant Numbers 19H05790, 20H00354 and 21H05233).

\section{ Appendix: Infrared radiation-induced photocurrents}
\label{appendix} 

\begin{figure}
	\centering
	\includegraphics[width=0.7\linewidth]{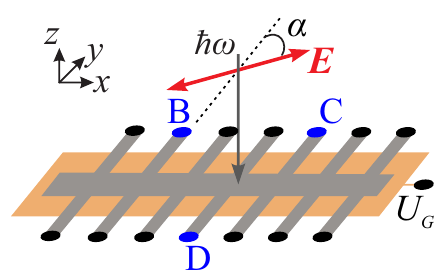}
	\caption{Sketch of the sample structure. The azimuth angle $\alpha$ is defined between the vector of the radiation electric field $\bm E$ and the short side of the Hall bar. Photosignals and transport were measured using the contacts marked in blue. 
	}
	\label{appendix_setup}
\end{figure}

\begin{figure}
	\centering
	\includegraphics[width=\linewidth]{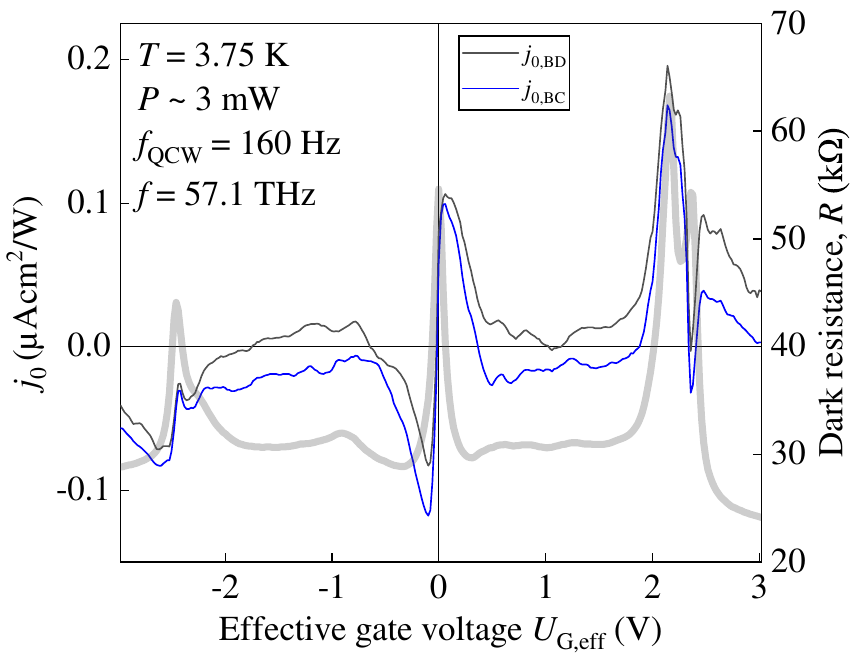}
	\caption{The gate voltage dependencies of polarization-independent component $j_0$ of the photocurrent, see Eq.~\eqref{parameters_definition}, measured using contact pairs BC and BD. Thick gray line: dark resistance measured between the contacts B and C.}
	\label{Appendix1}
\end{figure}

\begin{figure}
	\centering
	\includegraphics[width=\linewidth]{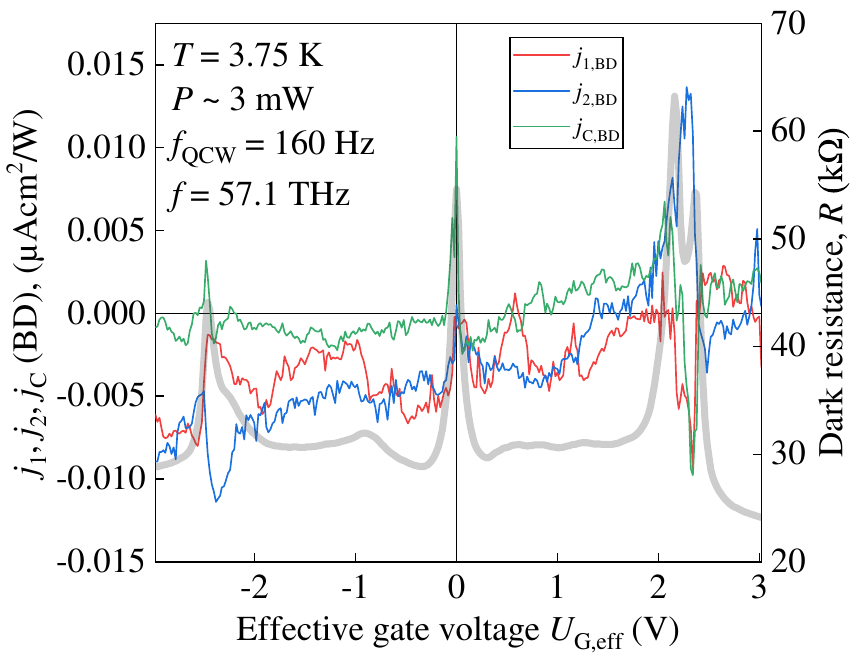}
	\caption{The gate voltage dependencies of polarization-sensitive photocurrent components $j_1$, $j_2$, and $j_\text{C}$, see Eq.~\eqref{parameters_definition} and \eqref{parameters_definition1}, measured between contacts B and D. Thick gray line: dark resistance measured between the contacts B and C.}
	\label{Appendix2}
\end{figure}

\begin{figure}
	\centering
	\includegraphics[width=\linewidth]{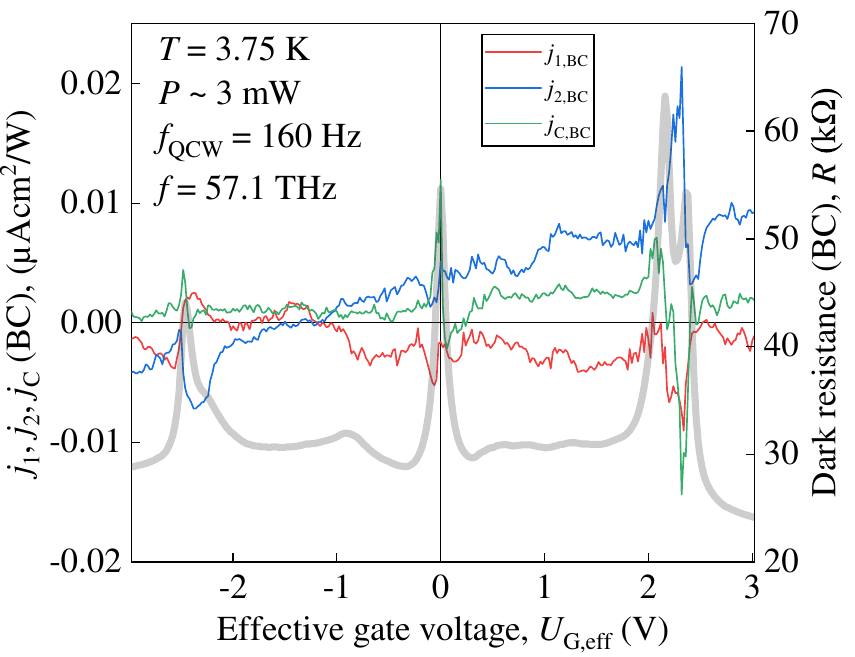}
	\caption{The gate voltage dependencies of polarization-sensitive photocurrent components $j_1$, $j_2$, and $j_\text{C}$, see Eq.~\eqref{parameters_definition} and \eqref{parameters_definition1}, measured between contacts B and C. Thick gray line: dark resistance measured between the contacts B and C.}
	\label{Appendix2BC}
\end{figure}

Apart from the photoresistance studied in a biased \gls{tblg}, in the absence of external bias we observed polarization-dependent photocurrents. Exemplary results, obtained at low $T=\SI{3.6}{\kelvin}$, are shown in Figs.~\ref{Appendix1}-\ref{Appendix2BC}. For the photocurrent measurements the continuous infrared \TeraHertz{57.1} radiation produced by \gls{qcl} was electronically modulated at a frequency of $f_{\text{QCW}}=\SI{160}{\hertz}$. The photocurrent along and across the Hall bar was measured using standard lock-in technique as a voltage drop between different pairs of contacts BC and BD, see Fig.~\ref{appendix_setup}. 

In the case of linear polarization and low intensity $I$, the photocurrent $j\propto I$ can be generally represented as \cite{Otteneder2020}
\begin{equation}
	j=j_{0}-j_{1}\cos(2\alpha)-j_{2}\sin(2\alpha)\,,
	\label{parameters_definition}
\end{equation}
where the components $j_0$, $j_1$, and $j_2$ are coefficients in front of the first three  Stokes parameters, and the azimuth angle $\alpha$ is defined as the angle of linear polarization with respect to $y$-direction across the Hall bar, see Fig.~\ref{appendix_setup}. The last linearly independent component of the photocurrent $j_\text{C}$, proportional to the forth Stokes parameter, requires application of the circularly polarized radiation, in which case the second and third Stokes parameters vanish, and 
\begin{equation}
	j=j_{0}+ j_\text{C} \eta\,.
	\label{parameters_definition1}
\end{equation}
The helicity-dependent photocurrent contribution $j_{\text{C}}\eta$, proportional to  the radiation helicity $\eta=\pm 1$, has opposite signs for the right- and left-handed circularly polarized radiation.

 \iffalse
 For the photocurrent measurements the \gls{qcl} was electronically modulated at a frequency of $f_{\text{QCW}}=\SI{160}{\hertz}$ and the current was measured as a voltage drop along the sample using standard lock-in technique. Note that there was no bias applied to the sample for the photocurrent measurements. The direction  parallel to the long Hall bar side was defined as the $x$-direction, while the perpendicular direction was defined as the $y$-direction, see Fig.~\ref{appendix_setup}. The angle between the radiation electric field and the $y$-direction is defined as the azimuth angle $\alpha$, see Fig.~\ref{appendix_setup}. Generally, the photocurrent varies with the azimuth angle $\alpha$ defining the rotation of linear polarization according to \cite{Otteneder2020}
\begin{equation}
	j=j_{0}-j_{1}\cos(2\alpha)-j_{2}\sin(2\alpha)\,,
	%\label{parameters_definition}
\end{equation}
which corresponds to a linear combination of the first three Stokes parameter weighted by the photocurrent coefficients $j_0$, $j_1$, and $j_2$.
The illumination of the \gls{tblg} structure with circularly polarized radiation also resulted in a helicity-dependent photocurrent contribution $j_{\text{C}}$. This photocurrent contribution reverses its sign if the radiation helicity is varied between right- and left-handed circularly polarized radiation and, consequently, corresponds to the last Stokes parameter.

\fi
Figure \ref{Appendix1} shows the extracted polarization-independent contribution of the photocurrent $j_0$ as a function of the applied gate voltage for both measurement directions BC and BD. It is seen that the photocurrent behaves similarly for both measurement directions. Similar to the photoresistance presented in the main text, the photocurrent shows pronounced features corresponding to peaks in the sample resistance (thick gray line in Fig.~\ref{Appendix1}). However, in contrast to the photoresistance, at the \gls{cnp} the photocurrent changes sign, together with the change of the charge of the majority carriers. Phenomenologically, these features are in line with results obtained earlier for \gls{tblg} in the terahertz range of radiation frequencies, see Refs.~\cite{Otteneder2020,Hubmann2022}. At the same time, the theory developed in Refs.~\cite{Otteneder2020} describes photocurrents induced via the intraband terahertz absorption, and thus is not directly applicable to the present case of direct interband transitions induced by the infrared illumination. Investigation of the corresponding mechanisms of sign-alternating \gls{tblg} photocurrents due to the interband absorption remains an interesting subject for future work.

Figures \ref{Appendix2} and \ref{Appendix2BC} show the gate voltage dependences of the extracted photocurrent components $j_1$ and $j_2$, sensitive to the direction of the linear polarization, as well as the helicity-sensitive contribution $j_{\text{C}}$, separately for the transverse (contacts BD, Fig.~\ref{Appendix2}) and longitudinal (contacts BC, Fig.~\ref{Appendix2BC}) photocurrents. Similar to the polarization-independent component $j_0$, the polarization-sensitive components possess similar features for both measurement directions correlated with the peaks in the sample resistance (grey lines). In contrast to the photocurrents detected in previous studies in the terahertz frequency range \cite{Otteneder2020,Hubmann2022}, in the infrared range the polarization-sensitive components are found to be much smaller than the polarization-independent photocurrent, see Fig.~\ref{Appendix1}.


%\FloatBarrier
%use standard all_bibfile if found, else use libfile in directory
\bibliography{all_lib}

\begin{thebibliography}{48}%
\makeatletter
\providecommand \@ifxundefined [1]{%
 \@ifx{#1\undefined}
}%
\providecommand \@ifnum [1]{%
 \ifnum #1\expandafter \@firstoftwo
 \else \expandafter \@secondoftwo
 \fi
}%
\providecommand \@ifx [1]{%
 \ifx #1\expandafter \@firstoftwo
 \else \expandafter \@secondoftwo
 \fi
}%
\providecommand \natexlab [1]{#1}%
\providecommand \enquote  [1]{``#1''}%
\providecommand \bibnamefont  [1]{#1}%
\providecommand \bibfnamefont [1]{#1}%
\providecommand \citenamefont [1]{#1}%
\providecommand \href@noop [0]{\@secondoftwo}%
\providecommand \href [0]{\begingroup \@sanitize@url \@href}%
\providecommand \@href[1]{\@@startlink{#1}\@@href}%
\providecommand \@@href[1]{\endgroup#1\@@endlink}%
\providecommand \@sanitize@url [0]{\catcode `\\12\catcode `\$12\catcode
  `\&12\catcode `\#12\catcode `\^12\catcode `\_12\catcode `\%12\relax}%
\providecommand \@@startlink[1]{}%
\providecommand \@@endlink[0]{}%
\providecommand \url  [0]{\begingroup\@sanitize@url \@url }%
\providecommand \@url [1]{\endgroup\@href {#1}{\urlprefix }}%
\providecommand \urlprefix  [0]{URL }%
\providecommand \Eprint [0]{\href }%
\providecommand \doibase [0]{http://dx.doi.org/}%
\providecommand \selectlanguage [0]{\@gobble}%
\providecommand \bibinfo  [0]{\@secondoftwo}%
\providecommand \bibfield  [0]{\@secondoftwo}%
\providecommand \translation [1]{[#1]}%
\providecommand \BibitemOpen [0]{}%
\providecommand \bibitemStop [0]{}%
\providecommand \bibitemNoStop [0]{.\EOS\space}%
\providecommand \EOS [0]{\spacefactor3000\relax}%
\providecommand \BibitemShut  [1]{\csname bibitem#1\endcsname}%
\let\auto@bib@innerbib\@empty
%</preamble>
\bibitem [{\citenamefont {Bistritzer}\ and\ \citenamefont
  {MacDonald}(2011)}]{Bistritzer2011}%
  \BibitemOpen
  \bibfield  {author} {\bibinfo {author} {\bibfnamefont {R.}~\bibnamefont
  {Bistritzer}}\ and\ \bibinfo {author} {\bibfnamefont {A.~H.}\ \bibnamefont
  {MacDonald}},\ }\href {\doibase 10.1073/pnas.1108174108} {\bibfield
  {journal} {\bibinfo  {journal} {Proc. Natl. Acad. Sci. U.S.A.}\ }\textbf
  {\bibinfo {volume} {108}},\ \bibinfo {pages} {12233} (\bibinfo {year}
  {2011})}\BibitemShut {NoStop}%
\bibitem [{\citenamefont {Cao}\ \emph {et~al.}(2018{\natexlab{a}})\citenamefont
  {Cao}, \citenamefont {Fatemi}, \citenamefont {Fang}, \citenamefont
  {Watanabe}, \citenamefont {Taniguchi}, \citenamefont {Kaxiras},\ and\
  \citenamefont {Jarillo-Herrero}}]{Cao2018}%
  \BibitemOpen
  \bibfield  {author} {\bibinfo {author} {\bibfnamefont {Y.}~\bibnamefont
  {Cao}}, \bibinfo {author} {\bibfnamefont {V.}~\bibnamefont {Fatemi}},
  \bibinfo {author} {\bibfnamefont {S.}~\bibnamefont {Fang}}, \bibinfo {author}
  {\bibfnamefont {K.}~\bibnamefont {Watanabe}}, \bibinfo {author}
  {\bibfnamefont {T.}~\bibnamefont {Taniguchi}}, \bibinfo {author}
  {\bibfnamefont {E.}~\bibnamefont {Kaxiras}}, \ and\ \bibinfo {author}
  {\bibfnamefont {P.}~\bibnamefont {Jarillo-Herrero}},\ }\href {\doibase
  10.1038/nature26160} {\bibfield  {journal} {\bibinfo  {journal} {Nature}\
  }\textbf {\bibinfo {volume} {556}},\ \bibinfo {pages} {43} (\bibinfo {year}
  {2018}{\natexlab{a}})}\BibitemShut {NoStop}%
\bibitem [{\citenamefont {Cao}\ \emph {et~al.}(2018{\natexlab{b}})\citenamefont
  {Cao}, \citenamefont {Fatemi}, \citenamefont {Demir}, \citenamefont {Fang},
  \citenamefont {Tomarken}, \citenamefont {Luo}, \citenamefont
  {Sanchez-Yamagishi}, \citenamefont {Watanabe}, \citenamefont {Taniguchi},
  \citenamefont {Kaxiras}, \citenamefont {Ashoori},\ and\ \citenamefont
  {Jarillo-Herrero}}]{Cao2018a}%
  \BibitemOpen
  \bibfield  {author} {\bibinfo {author} {\bibfnamefont {Y.}~\bibnamefont
  {Cao}}, \bibinfo {author} {\bibfnamefont {V.}~\bibnamefont {Fatemi}},
  \bibinfo {author} {\bibfnamefont {A.}~\bibnamefont {Demir}}, \bibinfo
  {author} {\bibfnamefont {S.}~\bibnamefont {Fang}}, \bibinfo {author}
  {\bibfnamefont {S.~L.}\ \bibnamefont {Tomarken}}, \bibinfo {author}
  {\bibfnamefont {J.~Y.}\ \bibnamefont {Luo}}, \bibinfo {author} {\bibfnamefont
  {J.~D.}\ \bibnamefont {Sanchez-Yamagishi}}, \bibinfo {author} {\bibfnamefont
  {K.}~\bibnamefont {Watanabe}}, \bibinfo {author} {\bibfnamefont
  {T.}~\bibnamefont {Taniguchi}}, \bibinfo {author} {\bibfnamefont
  {E.}~\bibnamefont {Kaxiras}}, \bibinfo {author} {\bibfnamefont {R.~C.}\
  \bibnamefont {Ashoori}}, \ and\ \bibinfo {author} {\bibfnamefont
  {P.}~\bibnamefont {Jarillo-Herrero}},\ }\href {\doibase 10.1038/nature26154}
  {\bibfield  {journal} {\bibinfo  {journal} {Nature}\ }\textbf {\bibinfo
  {volume} {556}},\ \bibinfo {pages} {80} (\bibinfo {year}
  {2018}{\natexlab{b}})}\BibitemShut {NoStop}%
\bibitem [{\citenamefont {Yankowitz}\ \emph {et~al.}(2019)\citenamefont
  {Yankowitz}, \citenamefont {Chen}, \citenamefont {Polshyn}, \citenamefont
  {Zhang}, \citenamefont {Watanabe}, \citenamefont {Taniguchi}, \citenamefont
  {Graf}, \citenamefont {Young},\ and\ \citenamefont {Dean}}]{Yankowitz2019}%
  \BibitemOpen
  \bibfield  {author} {\bibinfo {author} {\bibfnamefont {M.}~\bibnamefont
  {Yankowitz}}, \bibinfo {author} {\bibfnamefont {S.}~\bibnamefont {Chen}},
  \bibinfo {author} {\bibfnamefont {H.}~\bibnamefont {Polshyn}}, \bibinfo
  {author} {\bibfnamefont {Y.}~\bibnamefont {Zhang}}, \bibinfo {author}
  {\bibfnamefont {K.}~\bibnamefont {Watanabe}}, \bibinfo {author}
  {\bibfnamefont {T.}~\bibnamefont {Taniguchi}}, \bibinfo {author}
  {\bibfnamefont {D.}~\bibnamefont {Graf}}, \bibinfo {author} {\bibfnamefont
  {A.~F.}\ \bibnamefont {Young}}, \ and\ \bibinfo {author} {\bibfnamefont
  {C.~R.}\ \bibnamefont {Dean}},\ }\href {\doibase 10.1126/science.aav1910}
  {\bibfield  {journal} {\bibinfo  {journal} {Science}\ }\textbf {\bibinfo
  {volume} {363}},\ \bibinfo {pages} {1059} (\bibinfo {year}
  {2019})}\BibitemShut {NoStop}%
\bibitem [{\citenamefont {Sharpe}\ \emph {et~al.}(2019)\citenamefont {Sharpe},
  \citenamefont {Fox}, \citenamefont {Barnard}, \citenamefont {Finney},
  \citenamefont {Watanabe}, \citenamefont {Taniguchi}, \citenamefont
  {Kastner},\ and\ \citenamefont {Goldhaber-Gordon}}]{Sharpe2019}%
  \BibitemOpen
  \bibfield  {author} {\bibinfo {author} {\bibfnamefont {A.~L.}\ \bibnamefont
  {Sharpe}}, \bibinfo {author} {\bibfnamefont {E.~J.}\ \bibnamefont {Fox}},
  \bibinfo {author} {\bibfnamefont {A.~W.}\ \bibnamefont {Barnard}}, \bibinfo
  {author} {\bibfnamefont {J.}~\bibnamefont {Finney}}, \bibinfo {author}
  {\bibfnamefont {K.}~\bibnamefont {Watanabe}}, \bibinfo {author}
  {\bibfnamefont {T.}~\bibnamefont {Taniguchi}}, \bibinfo {author}
  {\bibfnamefont {M.~A.}\ \bibnamefont {Kastner}}, \ and\ \bibinfo {author}
  {\bibfnamefont {D.}~\bibnamefont {Goldhaber-Gordon}},\ }\href {\doibase
  10.1126/science.aaw3780} {\bibfield  {journal} {\bibinfo  {journal}
  {Science}\ }\textbf {\bibinfo {volume} {365}},\ \bibinfo {pages} {605}
  (\bibinfo {year} {2019})}\BibitemShut {NoStop}%
\bibitem [{\citenamefont {Serlin}\ \emph {et~al.}(2020)\citenamefont {Serlin},
  \citenamefont {Tschirhart}, \citenamefont {Polshyn}, \citenamefont {Zhang},
  \citenamefont {Zhu}, \citenamefont {Watanabe}, \citenamefont {Taniguchi},
  \citenamefont {Balents},\ and\ \citenamefont {Young}}]{Serlin2020}%
  \BibitemOpen
  \bibfield  {author} {\bibinfo {author} {\bibfnamefont {M.}~\bibnamefont
  {Serlin}}, \bibinfo {author} {\bibfnamefont {C.~L.}\ \bibnamefont
  {Tschirhart}}, \bibinfo {author} {\bibfnamefont {H.}~\bibnamefont {Polshyn}},
  \bibinfo {author} {\bibfnamefont {Y.}~\bibnamefont {Zhang}}, \bibinfo
  {author} {\bibfnamefont {J.}~\bibnamefont {Zhu}}, \bibinfo {author}
  {\bibfnamefont {K.}~\bibnamefont {Watanabe}}, \bibinfo {author}
  {\bibfnamefont {T.}~\bibnamefont {Taniguchi}}, \bibinfo {author}
  {\bibfnamefont {L.}~\bibnamefont {Balents}}, \ and\ \bibinfo {author}
  {\bibfnamefont {A.~F.}\ \bibnamefont {Young}},\ }\href {\doibase
  10.1126/science.aay5533} {\bibfield  {journal} {\bibinfo  {journal}
  {Science}\ }\textbf {\bibinfo {volume} {367}},\ \bibinfo {pages} {900}
  (\bibinfo {year} {2020})}\BibitemShut {NoStop}%
\bibitem [{\citenamefont {Andrei}\ and\ \citenamefont
  {MacDonald}(2020)}]{Andrei2020}%
  \BibitemOpen
  \bibfield  {author} {\bibinfo {author} {\bibfnamefont {E.~Y.}\ \bibnamefont
  {Andrei}}\ and\ \bibinfo {author} {\bibfnamefont {A.~H.}\ \bibnamefont
  {MacDonald}},\ }\href {\doibase 10.1038/s41563-020-00840-0} {\bibfield
  {journal} {\bibinfo  {journal} {Nat. Mater.}\ }\textbf {\bibinfo {volume}
  {19}},\ \bibinfo {pages} {1265} (\bibinfo {year} {2020})}\BibitemShut
  {NoStop}%
\bibitem [{\citenamefont {Törmä}\ \emph {et~al.}(2022)\citenamefont
  {Törmä}, \citenamefont {Peotta},\ and\ \citenamefont
  {Bernevig}}]{Toermae2022}%
  \BibitemOpen
  \bibfield  {author} {\bibinfo {author} {\bibfnamefont {P.}~\bibnamefont
  {Törmä}}, \bibinfo {author} {\bibfnamefont {S.}~\bibnamefont {Peotta}}, \
  and\ \bibinfo {author} {\bibfnamefont {B.~A.}\ \bibnamefont {Bernevig}},\
  }\href {\doibase 10.1038/s42254-022-00466-y} {\bibfield  {journal} {\bibinfo
  {journal} {Nat. Rev. Phys.}\ } (\bibinfo {year} {2022}),\
  10.1038/s42254-022-00466-y}\BibitemShut {NoStop}%
\bibitem [{\citenamefont {Song}\ and\ \citenamefont
  {Bernevig}(2022)}]{Song2022}%
  \BibitemOpen
  \bibfield  {author} {\bibinfo {author} {\bibfnamefont {Z.-D.}\ \bibnamefont
  {Song}}\ and\ \bibinfo {author} {\bibfnamefont {B.~A.}\ \bibnamefont
  {Bernevig}},\ }\href {\doibase 10.1103/PhysRevLett.129.047601} {\bibfield
  {journal} {\bibinfo  {journal} {Phys. Rev. Lett.}\ }\textbf {\bibinfo
  {volume} {129}},\ \bibinfo {pages} {047601} (\bibinfo {year}
  {2022})}\BibitemShut {NoStop}%
\bibitem [{\citenamefont {Chaudhary}\ \emph {et~al.}(2022)\citenamefont
  {Chaudhary}, \citenamefont {Lewandowski},\ and\ \citenamefont
  {Refael}}]{Chaudhary2022}%
  \BibitemOpen
  \bibfield  {author} {\bibinfo {author} {\bibfnamefont {S.}~\bibnamefont
  {Chaudhary}}, \bibinfo {author} {\bibfnamefont {C.}~\bibnamefont
  {Lewandowski}}, \ and\ \bibinfo {author} {\bibfnamefont {G.}~\bibnamefont
  {Refael}},\ }\href {\doibase 10.1103/PhysRevResearch.4.013164} {\bibfield
  {journal} {\bibinfo  {journal} {Phys. Rev. Research}\ }\textbf {\bibinfo
  {volume} {4}},\ \bibinfo {pages} {013164} (\bibinfo {year}
  {2022})}\BibitemShut {NoStop}%
\bibitem [{\citenamefont {Jaoui}\ \emph {et~al.}(2022)\citenamefont {Jaoui},
  \citenamefont {Das}, \citenamefont {Di~Battista}, \citenamefont
  {Díez-Mérida}, \citenamefont {Lu}, \citenamefont {Watanabe}, \citenamefont
  {Taniguchi}, \citenamefont {Ishizuka}, \citenamefont {Levitov},\ and\
  \citenamefont {Efetov}}]{Jaoui2022}%
  \BibitemOpen
  \bibfield  {author} {\bibinfo {author} {\bibfnamefont {A.}~\bibnamefont
  {Jaoui}}, \bibinfo {author} {\bibfnamefont {I.}~\bibnamefont {Das}}, \bibinfo
  {author} {\bibfnamefont {G.}~\bibnamefont {Di~Battista}}, \bibinfo {author}
  {\bibfnamefont {J.}~\bibnamefont {Díez-Mérida}}, \bibinfo {author}
  {\bibfnamefont {X.}~\bibnamefont {Lu}}, \bibinfo {author} {\bibfnamefont
  {K.}~\bibnamefont {Watanabe}}, \bibinfo {author} {\bibfnamefont
  {T.}~\bibnamefont {Taniguchi}}, \bibinfo {author} {\bibfnamefont
  {H.}~\bibnamefont {Ishizuka}}, \bibinfo {author} {\bibfnamefont
  {L.}~\bibnamefont {Levitov}}, \ and\ \bibinfo {author} {\bibfnamefont
  {D.~K.}\ \bibnamefont {Efetov}},\ }\href {\doibase
  10.1038/s41567-022-01556-5} {\bibfield  {journal} {\bibinfo  {journal} {Nat.
  Phys.}\ }\textbf {\bibinfo {volume} {18}},\ \bibinfo {pages} {633–638}
  (\bibinfo {year} {2022})}\BibitemShut {NoStop}%
\bibitem [{\citenamefont {Grover}\ \emph {et~al.}(2022)\citenamefont {Grover},
  \citenamefont {Bocarsly}, \citenamefont {Uri}, \citenamefont {Stepanov},
  \citenamefont {Battista}, \citenamefont {Roy}, \citenamefont {Xiao},
  \citenamefont {Meltzer}, \citenamefont {Myasoedov}, \citenamefont {Pareek},
  \citenamefont {Watanabe}, \citenamefont {Taniguchi}, \citenamefont {Yan},
  \citenamefont {Stern}, \citenamefont {Berg}, \citenamefont {Efetov},\ and\
  \citenamefont {Zeldov}}]{Grover2022}%
  \BibitemOpen
  \bibfield  {author} {\bibinfo {author} {\bibfnamefont {S.}~\bibnamefont
  {Grover}}, \bibinfo {author} {\bibfnamefont {M.}~\bibnamefont {Bocarsly}},
  \bibinfo {author} {\bibfnamefont {A.}~\bibnamefont {Uri}}, \bibinfo {author}
  {\bibfnamefont {P.}~\bibnamefont {Stepanov}}, \bibinfo {author}
  {\bibfnamefont {G.~D.}\ \bibnamefont {Battista}}, \bibinfo {author}
  {\bibfnamefont {I.}~\bibnamefont {Roy}}, \bibinfo {author} {\bibfnamefont
  {J.}~\bibnamefont {Xiao}}, \bibinfo {author} {\bibfnamefont {A.~Y.}\
  \bibnamefont {Meltzer}}, \bibinfo {author} {\bibfnamefont {Y.}~\bibnamefont
  {Myasoedov}}, \bibinfo {author} {\bibfnamefont {K.}~\bibnamefont {Pareek}},
  \bibinfo {author} {\bibfnamefont {K.}~\bibnamefont {Watanabe}}, \bibinfo
  {author} {\bibfnamefont {T.}~\bibnamefont {Taniguchi}}, \bibinfo {author}
  {\bibfnamefont {B.}~\bibnamefont {Yan}}, \bibinfo {author} {\bibfnamefont
  {A.}~\bibnamefont {Stern}}, \bibinfo {author} {\bibfnamefont
  {E.}~\bibnamefont {Berg}}, \bibinfo {author} {\bibfnamefont {D.~K.}\
  \bibnamefont {Efetov}}, \ and\ \bibinfo {author} {\bibfnamefont
  {E.}~\bibnamefont {Zeldov}},\ }\href {\doibase 10.1038/s41567-022-01635-7}
  {\bibfield  {journal} {\bibinfo  {journal} {Nat. Phys.}\ } (\bibinfo {year}
  {2022}),\ 10.1038/s41567-022-01635-7}\BibitemShut {NoStop}%
\bibitem [{\citenamefont {Cao}\ \emph {et~al.}(2021)\citenamefont {Cao},
  \citenamefont {Rodan-Legrain}, \citenamefont {Park}, \citenamefont {Yuan},
  \citenamefont {Watanabe}, \citenamefont {Taniguchi}, \citenamefont
  {Fernandes}, \citenamefont {Fu},\ and\ \citenamefont
  {Jarillo-Herrero}}]{Cao2021}%
  \BibitemOpen
  \bibfield  {author} {\bibinfo {author} {\bibfnamefont {Y.}~\bibnamefont
  {Cao}}, \bibinfo {author} {\bibfnamefont {D.}~\bibnamefont {Rodan-Legrain}},
  \bibinfo {author} {\bibfnamefont {J.~M.}\ \bibnamefont {Park}}, \bibinfo
  {author} {\bibfnamefont {N.~F.~Q.}\ \bibnamefont {Yuan}}, \bibinfo {author}
  {\bibfnamefont {K.}~\bibnamefont {Watanabe}}, \bibinfo {author}
  {\bibfnamefont {T.}~\bibnamefont {Taniguchi}}, \bibinfo {author}
  {\bibfnamefont {R.~M.}\ \bibnamefont {Fernandes}}, \bibinfo {author}
  {\bibfnamefont {L.}~\bibnamefont {Fu}}, \ and\ \bibinfo {author}
  {\bibfnamefont {P.}~\bibnamefont {Jarillo-Herrero}},\ }\href {\doibase
  10.1126/science.abc2836} {\bibfield  {journal} {\bibinfo  {journal}
  {Science}\ }\textbf {\bibinfo {volume} {372}},\ \bibinfo {pages} {264}
  (\bibinfo {year} {2021})}\BibitemShut {NoStop}%
\bibitem [{\citenamefont {Hao}\ \emph {et~al.}(2021)\citenamefont {Hao},
  \citenamefont {Zimmerman}, \citenamefont {Ledwith}, \citenamefont {Khalaf},
  \citenamefont {Najafabadi}, \citenamefont {Watanabe}, \citenamefont
  {Taniguchi}, \citenamefont {Vishwanath},\ and\ \citenamefont
  {Kim}}]{Hao2021}%
  \BibitemOpen
  \bibfield  {author} {\bibinfo {author} {\bibfnamefont {Z.}~\bibnamefont
  {Hao}}, \bibinfo {author} {\bibfnamefont {A.~M.}\ \bibnamefont {Zimmerman}},
  \bibinfo {author} {\bibfnamefont {P.}~\bibnamefont {Ledwith}}, \bibinfo
  {author} {\bibfnamefont {E.}~\bibnamefont {Khalaf}}, \bibinfo {author}
  {\bibfnamefont {D.~H.}\ \bibnamefont {Najafabadi}}, \bibinfo {author}
  {\bibfnamefont {K.}~\bibnamefont {Watanabe}}, \bibinfo {author}
  {\bibfnamefont {T.}~\bibnamefont {Taniguchi}}, \bibinfo {author}
  {\bibfnamefont {A.}~\bibnamefont {Vishwanath}}, \ and\ \bibinfo {author}
  {\bibfnamefont {P.}~\bibnamefont {Kim}},\ }\href {\doibase
  10.1126/science.abg0399} {\bibfield  {journal} {\bibinfo  {journal}
  {Science}\ }\textbf {\bibinfo {volume} {371}},\ \bibinfo {pages} {1133}
  (\bibinfo {year} {2021})}\BibitemShut {NoStop}%
\bibitem [{\citenamefont {Wu}\ \emph {et~al.}(2021)\citenamefont {Wu},
  \citenamefont {Zhang}, \citenamefont {Watanabe}, \citenamefont {Taniguchi},\
  and\ \citenamefont {Andrei}}]{Wu2021}%
  \BibitemOpen
  \bibfield  {author} {\bibinfo {author} {\bibfnamefont {S.}~\bibnamefont
  {Wu}}, \bibinfo {author} {\bibfnamefont {Z.}~\bibnamefont {Zhang}}, \bibinfo
  {author} {\bibfnamefont {K.}~\bibnamefont {Watanabe}}, \bibinfo {author}
  {\bibfnamefont {T.}~\bibnamefont {Taniguchi}}, \ and\ \bibinfo {author}
  {\bibfnamefont {E.~Y.}\ \bibnamefont {Andrei}},\ }\href {\doibase
  10.1038/s41563-020-00911-2} {\bibfield  {journal} {\bibinfo  {journal} {Nat.
  Mater.}\ }\textbf {\bibinfo {volume} {20}},\ \bibinfo {pages} {488} (\bibinfo
  {year} {2021})}\BibitemShut {NoStop}%
\bibitem [{\citenamefont {Andrei}\ \emph {et~al.}(2021)\citenamefont {Andrei},
  \citenamefont {Efetov}, \citenamefont {Jarillo-Herrero}, \citenamefont
  {MacDonald}, \citenamefont {Mak}, \citenamefont {Senthil}, \citenamefont
  {Tutuc}, \citenamefont {Yazdani},\ and\ \citenamefont {Young}}]{Andrei2021}%
  \BibitemOpen
  \bibfield  {author} {\bibinfo {author} {\bibfnamefont {E.~Y.}\ \bibnamefont
  {Andrei}}, \bibinfo {author} {\bibfnamefont {D.~K.}\ \bibnamefont {Efetov}},
  \bibinfo {author} {\bibfnamefont {P.}~\bibnamefont {Jarillo-Herrero}},
  \bibinfo {author} {\bibfnamefont {A.~H.}\ \bibnamefont {MacDonald}}, \bibinfo
  {author} {\bibfnamefont {K.~F.}\ \bibnamefont {Mak}}, \bibinfo {author}
  {\bibfnamefont {T.}~\bibnamefont {Senthil}}, \bibinfo {author} {\bibfnamefont
  {E.}~\bibnamefont {Tutuc}}, \bibinfo {author} {\bibfnamefont
  {A.}~\bibnamefont {Yazdani}}, \ and\ \bibinfo {author} {\bibfnamefont
  {A.~F.}\ \bibnamefont {Young}},\ }\href {\doibase 10.1038/s41578-021-00284-1}
  {\bibfield  {journal} {\bibinfo  {journal} {Nat. Rev. Mater.}\ }\textbf
  {\bibinfo {volume} {6}},\ \bibinfo {pages} {201–206} (\bibinfo {year}
  {2021})}\BibitemShut {NoStop}%
\bibitem [{\citenamefont {Hesp}\ \emph
  {et~al.}(2021{\natexlab{a}})\citenamefont {Hesp}, \citenamefont {Torre},
  \citenamefont {Rodan-Legrain}, \citenamefont {Novelli}, \citenamefont {Cao},
  \citenamefont {Carr}, \citenamefont {Fang}, \citenamefont {Stepanov},
  \citenamefont {Barcons-Ruiz}, \citenamefont {Herzig-Sheinfux}, \citenamefont
  {Watanabe}, \citenamefont {Taniguchi}, \citenamefont {Efetov}, \citenamefont
  {Kaxiras}, \citenamefont {Jarillo-Herrero}, \citenamefont {Polini},\ and\
  \citenamefont {Koppens}}]{Hesp2021a}%
  \BibitemOpen
  \bibfield  {author} {\bibinfo {author} {\bibfnamefont {N.~C.~H.}\
  \bibnamefont {Hesp}}, \bibinfo {author} {\bibfnamefont {I.}~\bibnamefont
  {Torre}}, \bibinfo {author} {\bibfnamefont {D.}~\bibnamefont
  {Rodan-Legrain}}, \bibinfo {author} {\bibfnamefont {P.}~\bibnamefont
  {Novelli}}, \bibinfo {author} {\bibfnamefont {Y.}~\bibnamefont {Cao}},
  \bibinfo {author} {\bibfnamefont {S.}~\bibnamefont {Carr}}, \bibinfo {author}
  {\bibfnamefont {S.}~\bibnamefont {Fang}}, \bibinfo {author} {\bibfnamefont
  {P.}~\bibnamefont {Stepanov}}, \bibinfo {author} {\bibfnamefont
  {D.}~\bibnamefont {Barcons-Ruiz}}, \bibinfo {author} {\bibfnamefont
  {H.}~\bibnamefont {Herzig-Sheinfux}}, \bibinfo {author} {\bibfnamefont
  {K.}~\bibnamefont {Watanabe}}, \bibinfo {author} {\bibfnamefont
  {T.}~\bibnamefont {Taniguchi}}, \bibinfo {author} {\bibfnamefont {D.~K.}\
  \bibnamefont {Efetov}}, \bibinfo {author} {\bibfnamefont {E.}~\bibnamefont
  {Kaxiras}}, \bibinfo {author} {\bibfnamefont {P.}~\bibnamefont
  {Jarillo-Herrero}}, \bibinfo {author} {\bibfnamefont {M.}~\bibnamefont
  {Polini}}, \ and\ \bibinfo {author} {\bibfnamefont {F.~H.~L.}\ \bibnamefont
  {Koppens}},\ }\href {\doibase 10.1038/s41567-021-01327-8} {\bibfield
  {journal} {\bibinfo  {journal} {Nat. Phys.}\ }\textbf {\bibinfo {volume}
  {17}},\ \bibinfo {pages} {1162} (\bibinfo {year}
  {2021}{\natexlab{a}})}\BibitemShut {NoStop}%
\bibitem [{\citenamefont {Liu}\ \emph {et~al.}(2020)\citenamefont {Liu},
  \citenamefont {Hao}, \citenamefont {Khalaf}, \citenamefont {Lee},
  \citenamefont {Ronen}, \citenamefont {Yoo}, \citenamefont {Najafabadi},
  \citenamefont {Watanabe}, \citenamefont {Taniguchi}, \citenamefont
  {Vishwanath},\ and\ \citenamefont {Kim}}]{Liu2020}%
  \BibitemOpen
  \bibfield  {author} {\bibinfo {author} {\bibfnamefont {X.}~\bibnamefont
  {Liu}}, \bibinfo {author} {\bibfnamefont {Z.}~\bibnamefont {Hao}}, \bibinfo
  {author} {\bibfnamefont {E.}~\bibnamefont {Khalaf}}, \bibinfo {author}
  {\bibfnamefont {J.~Y.}\ \bibnamefont {Lee}}, \bibinfo {author} {\bibfnamefont
  {Y.}~\bibnamefont {Ronen}}, \bibinfo {author} {\bibfnamefont
  {H.}~\bibnamefont {Yoo}}, \bibinfo {author} {\bibfnamefont {D.~H.}\
  \bibnamefont {Najafabadi}}, \bibinfo {author} {\bibfnamefont
  {K.}~\bibnamefont {Watanabe}}, \bibinfo {author} {\bibfnamefont
  {T.}~\bibnamefont {Taniguchi}}, \bibinfo {author} {\bibfnamefont
  {A.}~\bibnamefont {Vishwanath}}, \ and\ \bibinfo {author} {\bibfnamefont
  {P.}~\bibnamefont {Kim}},\ }\href {\doibase 10.1038/s41586-020-2458-7}
  {\bibfield  {journal} {\bibinfo  {journal} {Nature}\ }\textbf {\bibinfo
  {volume} {583}},\ \bibinfo {pages} {221} (\bibinfo {year}
  {2020})}\BibitemShut {NoStop}%
\bibitem [{\citenamefont {Uri}\ \emph {et~al.}(2020)\citenamefont {Uri},
  \citenamefont {Grover}, \citenamefont {Cao}, \citenamefont {Crosse},
  \citenamefont {Bagani}, \citenamefont {Rodan-Legrain}, \citenamefont
  {Myasoedov}, \citenamefont {Watanabe}, \citenamefont {Taniguchi},
  \citenamefont {Moon},\ and\ \citenamefont {et~al.}}]{uri2020}%
  \BibitemOpen
  \bibfield  {author} {\bibinfo {author} {\bibfnamefont {A.}~\bibnamefont
  {Uri}}, \bibinfo {author} {\bibfnamefont {S.}~\bibnamefont {Grover}},
  \bibinfo {author} {\bibfnamefont {Y.}~\bibnamefont {Cao}}, \bibinfo {author}
  {\bibfnamefont {J.}~\bibnamefont {Crosse}}, \bibinfo {author} {\bibfnamefont
  {K.}~\bibnamefont {Bagani}}, \bibinfo {author} {\bibfnamefont
  {D.}~\bibnamefont {Rodan-Legrain}}, \bibinfo {author} {\bibfnamefont
  {Y.}~\bibnamefont {Myasoedov}}, \bibinfo {author} {\bibfnamefont
  {K.}~\bibnamefont {Watanabe}}, \bibinfo {author} {\bibfnamefont
  {T.}~\bibnamefont {Taniguchi}}, \bibinfo {author} {\bibfnamefont
  {P.}~\bibnamefont {Moon}}, \ and\ \bibinfo {author} {\bibnamefont {et~al.}},\
  }\href {\doibase 10.1038/s41586-020-2255-3} {\bibfield  {journal} {\bibinfo
  {journal} {Nature}\ }\textbf {\bibinfo {volume} {581}},\ \bibinfo {pages}
  {47–52} (\bibinfo {year} {2020})}\BibitemShut {NoStop}%
\bibitem [{\citenamefont {Cao}\ \emph {et~al.}(2020)\citenamefont {Cao},
  \citenamefont {Chowdhury}, \citenamefont {Rodan-Legrain}, \citenamefont
  {Rubies-Bigorda}, \citenamefont {Watanabe}, \citenamefont {Taniguchi},
  \citenamefont {Senthil},\ and\ \citenamefont {Jarillo-Herrero}}]{Cao2020}%
  \BibitemOpen
  \bibfield  {author} {\bibinfo {author} {\bibfnamefont {Y.}~\bibnamefont
  {Cao}}, \bibinfo {author} {\bibfnamefont {D.}~\bibnamefont {Chowdhury}},
  \bibinfo {author} {\bibfnamefont {D.}~\bibnamefont {Rodan-Legrain}}, \bibinfo
  {author} {\bibfnamefont {O.}~\bibnamefont {Rubies-Bigorda}}, \bibinfo
  {author} {\bibfnamefont {K.}~\bibnamefont {Watanabe}}, \bibinfo {author}
  {\bibfnamefont {T.}~\bibnamefont {Taniguchi}}, \bibinfo {author}
  {\bibfnamefont {T.}~\bibnamefont {Senthil}}, \ and\ \bibinfo {author}
  {\bibfnamefont {P.}~\bibnamefont {Jarillo-Herrero}},\ }\href {\doibase
  10.1103/physrevlett.124.076801} {\bibfield  {journal} {\bibinfo  {journal}
  {Phys. Rev. Lett.}\ }\textbf {\bibinfo {volume} {124}},\ \bibinfo {pages}
  {076801} (\bibinfo {year} {2020})}\BibitemShut {NoStop}%
\bibitem [{\citenamefont {Balents}\ \emph {et~al.}(2020)\citenamefont
  {Balents}, \citenamefont {Dean}, \citenamefont {Efetov},\ and\ \citenamefont
  {Young}}]{Balents2020}%
  \BibitemOpen
  \bibfield  {author} {\bibinfo {author} {\bibfnamefont {L.}~\bibnamefont
  {Balents}}, \bibinfo {author} {\bibfnamefont {C.~R.}\ \bibnamefont {Dean}},
  \bibinfo {author} {\bibfnamefont {D.~K.}\ \bibnamefont {Efetov}}, \ and\
  \bibinfo {author} {\bibfnamefont {A.~F.}\ \bibnamefont {Young}},\ }\href
  {\doibase 10.1038/s41567-020-0906-9} {\bibfield  {journal} {\bibinfo
  {journal} {Nat. Phys.}\ }\textbf {\bibinfo {volume} {16}},\ \bibinfo {pages}
  {725–733} (\bibinfo {year} {2020})}\BibitemShut {NoStop}%
\bibitem [{\citenamefont {Otteneder}\ \emph {et~al.}(2020)\citenamefont
  {Otteneder}, \citenamefont {Hubmann}, \citenamefont {Lu}, \citenamefont
  {Kozlov}, \citenamefont {Golub}, \citenamefont {Watanabe}, \citenamefont
  {Taniguchi}, \citenamefont {Efetov},\ and\ \citenamefont
  {Ganichev}}]{Otteneder2020}%
  \BibitemOpen
  \bibfield  {author} {\bibinfo {author} {\bibfnamefont {M.}~\bibnamefont
  {Otteneder}}, \bibinfo {author} {\bibfnamefont {S.}~\bibnamefont {Hubmann}},
  \bibinfo {author} {\bibfnamefont {X.}~\bibnamefont {Lu}}, \bibinfo {author}
  {\bibfnamefont {D.~A.}\ \bibnamefont {Kozlov}}, \bibinfo {author}
  {\bibfnamefont {L.~E.}\ \bibnamefont {Golub}}, \bibinfo {author}
  {\bibfnamefont {K.}~\bibnamefont {Watanabe}}, \bibinfo {author}
  {\bibfnamefont {T.}~\bibnamefont {Taniguchi}}, \bibinfo {author}
  {\bibfnamefont {D.~K.}\ \bibnamefont {Efetov}}, \ and\ \bibinfo {author}
  {\bibfnamefont {S.~D.}\ \bibnamefont {Ganichev}},\ }\href {\doibase
  10.1021/acs.nanolett.0c02474} {\bibfield  {journal} {\bibinfo  {journal}
  {Nano Lett.}\ }\textbf {\bibinfo {volume} {20}},\ \bibinfo {pages} {7152}
  (\bibinfo {year} {2020})}\BibitemShut {NoStop}%
\bibitem [{\citenamefont {Yoo}\ \emph {et~al.}(2019)\citenamefont {Yoo},
  \citenamefont {Engelke}, \citenamefont {Carr}, \citenamefont {Fang},
  \citenamefont {Zhang}, \citenamefont {Cazeaux}, \citenamefont {Sung},
  \citenamefont {Hovden}, \citenamefont {Tsen}, \citenamefont {Taniguchi},
  \citenamefont {Watanabe}, \citenamefont {Yi}, \citenamefont {Kim},
  \citenamefont {Luskin}, \citenamefont {Tadmor}, \citenamefont {Kaxiras},\
  and\ \citenamefont {Kim}}]{Yoo2019}%
  \BibitemOpen
  \bibfield  {author} {\bibinfo {author} {\bibfnamefont {H.}~\bibnamefont
  {Yoo}}, \bibinfo {author} {\bibfnamefont {R.}~\bibnamefont {Engelke}},
  \bibinfo {author} {\bibfnamefont {S.}~\bibnamefont {Carr}}, \bibinfo {author}
  {\bibfnamefont {S.}~\bibnamefont {Fang}}, \bibinfo {author} {\bibfnamefont
  {K.}~\bibnamefont {Zhang}}, \bibinfo {author} {\bibfnamefont
  {P.}~\bibnamefont {Cazeaux}}, \bibinfo {author} {\bibfnamefont {S.~H.}\
  \bibnamefont {Sung}}, \bibinfo {author} {\bibfnamefont {R.}~\bibnamefont
  {Hovden}}, \bibinfo {author} {\bibfnamefont {A.~W.}\ \bibnamefont {Tsen}},
  \bibinfo {author} {\bibfnamefont {T.}~\bibnamefont {Taniguchi}}, \bibinfo
  {author} {\bibfnamefont {K.}~\bibnamefont {Watanabe}}, \bibinfo {author}
  {\bibfnamefont {G.-C.}\ \bibnamefont {Yi}}, \bibinfo {author} {\bibfnamefont
  {M.}~\bibnamefont {Kim}}, \bibinfo {author} {\bibfnamefont {M.}~\bibnamefont
  {Luskin}}, \bibinfo {author} {\bibfnamefont {E.~B.}\ \bibnamefont {Tadmor}},
  \bibinfo {author} {\bibfnamefont {E.}~\bibnamefont {Kaxiras}}, \ and\
  \bibinfo {author} {\bibfnamefont {P.}~\bibnamefont {Kim}},\ }\href {\doibase
  10.1038/s41563-019-0346-z} {\bibfield  {journal} {\bibinfo  {journal} {Nat.
  Mater.}\ }\textbf {\bibinfo {volume} {18}},\ \bibinfo {pages} {448} (\bibinfo
  {year} {2019})}\BibitemShut {NoStop}%
\bibitem [{\citenamefont {Choi}\ \emph {et~al.}(2019)\citenamefont {Choi},
  \citenamefont {Kemmer}, \citenamefont {Peng}, \citenamefont {Thomson},
  \citenamefont {Arora}, \citenamefont {Polski}, \citenamefont {Zhang},
  \citenamefont {Ren}, \citenamefont {Alicea}, \citenamefont {Refael},
  \citenamefont {von Oppen}, \citenamefont {Watanabe}, \citenamefont
  {Taniguchi},\ and\ \citenamefont {Nadj-Perge}}]{Choi2019}%
  \BibitemOpen
  \bibfield  {author} {\bibinfo {author} {\bibfnamefont {Y.}~\bibnamefont
  {Choi}}, \bibinfo {author} {\bibfnamefont {J.}~\bibnamefont {Kemmer}},
  \bibinfo {author} {\bibfnamefont {Y.}~\bibnamefont {Peng}}, \bibinfo {author}
  {\bibfnamefont {A.}~\bibnamefont {Thomson}}, \bibinfo {author} {\bibfnamefont
  {H.}~\bibnamefont {Arora}}, \bibinfo {author} {\bibfnamefont
  {R.}~\bibnamefont {Polski}}, \bibinfo {author} {\bibfnamefont
  {Y.}~\bibnamefont {Zhang}}, \bibinfo {author} {\bibfnamefont
  {H.}~\bibnamefont {Ren}}, \bibinfo {author} {\bibfnamefont {J.}~\bibnamefont
  {Alicea}}, \bibinfo {author} {\bibfnamefont {G.}~\bibnamefont {Refael}},
  \bibinfo {author} {\bibfnamefont {F.}~\bibnamefont {von Oppen}}, \bibinfo
  {author} {\bibfnamefont {K.}~\bibnamefont {Watanabe}}, \bibinfo {author}
  {\bibfnamefont {T.}~\bibnamefont {Taniguchi}}, \ and\ \bibinfo {author}
  {\bibfnamefont {S.}~\bibnamefont {Nadj-Perge}},\ }\href {\doibase
  10.1038/s41567-019-0606-5} {\bibfield  {journal} {\bibinfo  {journal} {Nat.
  Phys.}\ }\textbf {\bibinfo {volume} {15}},\ \bibinfo {pages} {1174} (\bibinfo
  {year} {2019})}\BibitemShut {NoStop}%
\bibitem [{\citenamefont {Lu}\ \emph {et~al.}(2019)\citenamefont {Lu},
  \citenamefont {Stepanov}, \citenamefont {Yang}, \citenamefont {Xie},
  \citenamefont {Aamir}, \citenamefont {Das}, \citenamefont {Urgell},
  \citenamefont {Watanabe}, \citenamefont {Taniguchi}, \citenamefont {Zhang},
  \citenamefont {Bachtold}, \citenamefont {MacDonald},\ and\ \citenamefont
  {Efetov}}]{Lu2019}%
  \BibitemOpen
  \bibfield  {author} {\bibinfo {author} {\bibfnamefont {X.}~\bibnamefont
  {Lu}}, \bibinfo {author} {\bibfnamefont {P.}~\bibnamefont {Stepanov}},
  \bibinfo {author} {\bibfnamefont {W.}~\bibnamefont {Yang}}, \bibinfo {author}
  {\bibfnamefont {M.}~\bibnamefont {Xie}}, \bibinfo {author} {\bibfnamefont
  {M.~A.}\ \bibnamefont {Aamir}}, \bibinfo {author} {\bibfnamefont
  {I.}~\bibnamefont {Das}}, \bibinfo {author} {\bibfnamefont {C.}~\bibnamefont
  {Urgell}}, \bibinfo {author} {\bibfnamefont {K.}~\bibnamefont {Watanabe}},
  \bibinfo {author} {\bibfnamefont {T.}~\bibnamefont {Taniguchi}}, \bibinfo
  {author} {\bibfnamefont {G.}~\bibnamefont {Zhang}}, \bibinfo {author}
  {\bibfnamefont {A.}~\bibnamefont {Bachtold}}, \bibinfo {author}
  {\bibfnamefont {A.~H.}\ \bibnamefont {MacDonald}}, \ and\ \bibinfo {author}
  {\bibfnamefont {D.~K.}\ \bibnamefont {Efetov}},\ }\href {\doibase
  10.1038/s41586-019-1695-0} {\bibfield  {journal} {\bibinfo  {journal}
  {Nature}\ }\textbf {\bibinfo {volume} {574}},\ \bibinfo {pages} {653}
  (\bibinfo {year} {2019})}\BibitemShut {NoStop}%
\bibitem [{\citenamefont {Seifert}\ \emph {et~al.}(2020)\citenamefont
  {Seifert}, \citenamefont {Lu}, \citenamefont {Stepanov}, \citenamefont
  {Durán~Retamal}, \citenamefont {Moore}, \citenamefont {Fong}, \citenamefont
  {Principi},\ and\ \citenamefont {Efetov}}]{Seifert2020}%
  \BibitemOpen
  \bibfield  {author} {\bibinfo {author} {\bibfnamefont {P.}~\bibnamefont
  {Seifert}}, \bibinfo {author} {\bibfnamefont {X.}~\bibnamefont {Lu}},
  \bibinfo {author} {\bibfnamefont {P.}~\bibnamefont {Stepanov}}, \bibinfo
  {author} {\bibfnamefont {J.~R.}\ \bibnamefont {Durán~Retamal}}, \bibinfo
  {author} {\bibfnamefont {J.~N.}\ \bibnamefont {Moore}}, \bibinfo {author}
  {\bibfnamefont {K.-C.}\ \bibnamefont {Fong}}, \bibinfo {author}
  {\bibfnamefont {A.}~\bibnamefont {Principi}}, \ and\ \bibinfo {author}
  {\bibfnamefont {D.~K.}\ \bibnamefont {Efetov}},\ }\href {\doibase
  10.1021/acs.nanolett.0c00373} {\bibfield  {journal} {\bibinfo  {journal}
  {Nano Lett.}\ }\textbf {\bibinfo {volume} {20}},\ \bibinfo {pages} {3459}
  (\bibinfo {year} {2020})}\BibitemShut {NoStop}%
\bibitem [{\citenamefont {Di~Battista}\ \emph {et~al.}(2021)\citenamefont
  {Di~Battista}, \citenamefont {Seifert}, \citenamefont {Watanabe},
  \citenamefont {Taniguchi}, \citenamefont {Fong}, \citenamefont {Principi},\
  and\ \citenamefont {Efetov}}]{DiBattista2021}%
  \BibitemOpen
  \bibfield  {author} {\bibinfo {author} {\bibfnamefont {G.}~\bibnamefont
  {Di~Battista}}, \bibinfo {author} {\bibfnamefont {P.}~\bibnamefont
  {Seifert}}, \bibinfo {author} {\bibfnamefont {K.}~\bibnamefont {Watanabe}},
  \bibinfo {author} {\bibfnamefont {T.}~\bibnamefont {Taniguchi}}, \bibinfo
  {author} {\bibfnamefont {K.}~\bibnamefont {Fong}}, \bibinfo {author}
  {\bibfnamefont {A.}~\bibnamefont {Principi}}, \ and\ \bibinfo {author}
  {\bibfnamefont {D.}~\bibnamefont {Efetov}},\ }\href@noop {} {\bibfield
  {journal} {\bibinfo  {journal} {arXiv preprint arXiv:2111.08735}\ } (\bibinfo
  {year} {2021})}\BibitemShut {NoStop}%
\bibitem [{\citenamefont {Xin}\ \emph {et~al.}(2016)\citenamefont {Xin},
  \citenamefont {Chen}, \citenamefont {Liu}, \citenamefont {Jiang},
  \citenamefont {Gao}, \citenamefont {Jiang}, \citenamefont {Chen},\ and\
  \citenamefont {Tian}}]{Xin2016}%
  \BibitemOpen
  \bibfield  {author} {\bibinfo {author} {\bibfnamefont {W.}~\bibnamefont
  {Xin}}, \bibinfo {author} {\bibfnamefont {X.-D.}\ \bibnamefont {Chen}},
  \bibinfo {author} {\bibfnamefont {Z.-B.}\ \bibnamefont {Liu}}, \bibinfo
  {author} {\bibfnamefont {W.-S.}\ \bibnamefont {Jiang}}, \bibinfo {author}
  {\bibfnamefont {X.-G.}\ \bibnamefont {Gao}}, \bibinfo {author} {\bibfnamefont
  {X.-Q.}\ \bibnamefont {Jiang}}, \bibinfo {author} {\bibfnamefont
  {Y.}~\bibnamefont {Chen}}, \ and\ \bibinfo {author} {\bibfnamefont {J.-G.}\
  \bibnamefont {Tian}},\ }\href {\doibase 10.1002/adom.201600278} {\bibfield
  {journal} {\bibinfo  {journal} {Adv. Opt. Mater.}\ }\textbf {\bibinfo
  {volume} {4}},\ \bibinfo {pages} {1703} (\bibinfo {year} {2016})}\BibitemShut
  {NoStop}%
\bibitem [{\citenamefont {Yin}\ \emph {et~al.}(2016)\citenamefont {Yin},
  \citenamefont {Wang}, \citenamefont {Peng}, \citenamefont {Tan},
  \citenamefont {Liao}, \citenamefont {Lin}, \citenamefont {Sun}, \citenamefont
  {Koh}, \citenamefont {Chen}, \citenamefont {Peng},\ and\ \citenamefont
  {Liu}}]{Yin2016a}%
  \BibitemOpen
  \bibfield  {author} {\bibinfo {author} {\bibfnamefont {J.}~\bibnamefont
  {Yin}}, \bibinfo {author} {\bibfnamefont {H.}~\bibnamefont {Wang}}, \bibinfo
  {author} {\bibfnamefont {H.}~\bibnamefont {Peng}}, \bibinfo {author}
  {\bibfnamefont {Z.}~\bibnamefont {Tan}}, \bibinfo {author} {\bibfnamefont
  {L.}~\bibnamefont {Liao}}, \bibinfo {author} {\bibfnamefont {L.}~\bibnamefont
  {Lin}}, \bibinfo {author} {\bibfnamefont {X.}~\bibnamefont {Sun}}, \bibinfo
  {author} {\bibfnamefont {A.~L.}\ \bibnamefont {Koh}}, \bibinfo {author}
  {\bibfnamefont {Y.}~\bibnamefont {Chen}}, \bibinfo {author} {\bibfnamefont
  {H.}~\bibnamefont {Peng}}, \ and\ \bibinfo {author} {\bibfnamefont
  {Z.}~\bibnamefont {Liu}},\ }\href {\doibase 10.1038/ncomms10699} {\bibfield
  {journal} {\bibinfo  {journal} {Nat. Comm.}\ }\textbf {\bibinfo {volume}
  {7}},\ \bibinfo {pages} {10699} (\bibinfo {year} {2016})}\BibitemShut
  {NoStop}%
\bibitem [{\citenamefont {Sunku}\ \emph {et~al.}(2020)\citenamefont {Sunku},
  \citenamefont {McLeod}, \citenamefont {Stauber}, \citenamefont {Yoo},
  \citenamefont {Halbertal}, \citenamefont {Ni}, \citenamefont {Sternbach},
  \citenamefont {Jiang}, \citenamefont {Taniguchi}, \citenamefont {Watanabe},
  \citenamefont {Kim}, \citenamefont {Fogler},\ and\ \citenamefont
  {Basov}}]{Sunku2020}%
  \BibitemOpen
  \bibfield  {author} {\bibinfo {author} {\bibfnamefont {S.~S.}\ \bibnamefont
  {Sunku}}, \bibinfo {author} {\bibfnamefont {A.~S.}\ \bibnamefont {McLeod}},
  \bibinfo {author} {\bibfnamefont {T.}~\bibnamefont {Stauber}}, \bibinfo
  {author} {\bibfnamefont {H.}~\bibnamefont {Yoo}}, \bibinfo {author}
  {\bibfnamefont {D.}~\bibnamefont {Halbertal}}, \bibinfo {author}
  {\bibfnamefont {G.}~\bibnamefont {Ni}}, \bibinfo {author} {\bibfnamefont
  {A.}~\bibnamefont {Sternbach}}, \bibinfo {author} {\bibfnamefont {B.-Y.}\
  \bibnamefont {Jiang}}, \bibinfo {author} {\bibfnamefont {T.}~\bibnamefont
  {Taniguchi}}, \bibinfo {author} {\bibfnamefont {K.}~\bibnamefont {Watanabe}},
  \bibinfo {author} {\bibfnamefont {P.}~\bibnamefont {Kim}}, \bibinfo {author}
  {\bibfnamefont {M.~M.}\ \bibnamefont {Fogler}}, \ and\ \bibinfo {author}
  {\bibfnamefont {D.~N.}\ \bibnamefont {Basov}},\ }\href {\doibase
  10.1021/acs.nanolett.9b04637} {\bibfield  {journal} {\bibinfo  {journal}
  {Nano Lett.}\ }\textbf {\bibinfo {volume} {20}},\ \bibinfo {pages} {2958}
  (\bibinfo {year} {2020})}\BibitemShut {NoStop}%
\bibitem [{\citenamefont {Hesp}\ \emph
  {et~al.}(2021{\natexlab{b}})\citenamefont {Hesp}, \citenamefont {Torre},
  \citenamefont {Barcons-Ruiz}, \citenamefont {Sheinfux}, \citenamefont
  {Watanabe}, \citenamefont {Taniguchi}, \citenamefont {Kumar},\ and\
  \citenamefont {Koppens}}]{Hesp2021}%
  \BibitemOpen
  \bibfield  {author} {\bibinfo {author} {\bibfnamefont {N.~C.~H.}\
  \bibnamefont {Hesp}}, \bibinfo {author} {\bibfnamefont {I.}~\bibnamefont
  {Torre}}, \bibinfo {author} {\bibfnamefont {D.}~\bibnamefont {Barcons-Ruiz}},
  \bibinfo {author} {\bibfnamefont {H.~H.}\ \bibnamefont {Sheinfux}}, \bibinfo
  {author} {\bibfnamefont {K.}~\bibnamefont {Watanabe}}, \bibinfo {author}
  {\bibfnamefont {T.}~\bibnamefont {Taniguchi}}, \bibinfo {author}
  {\bibfnamefont {R.~K.}\ \bibnamefont {Kumar}}, \ and\ \bibinfo {author}
  {\bibfnamefont {F.~H.~L.}\ \bibnamefont {Koppens}},\ }\href {\doibase
  10.1038/s41467-021-21862-5} {\bibfield  {journal} {\bibinfo  {journal} {Nat.
  Commun.}\ }\textbf {\bibinfo {volume} {12}},\ \bibinfo {pages} {1640}
  (\bibinfo {year} {2021}{\natexlab{b}})}\BibitemShut {NoStop}%
\bibitem [{\citenamefont {Sunku}\ \emph {et~al.}(2021)\citenamefont {Sunku},
  \citenamefont {Halbertal}, \citenamefont {Stauber}, \citenamefont {Chen},
  \citenamefont {McLeod}, \citenamefont {Rikhter}, \citenamefont {Berkowitz},
  \citenamefont {Lo}, \citenamefont {Gonzalez-Acevedo}, \citenamefont {Hone},
  \citenamefont {Dean}, \citenamefont {Fogler},\ and\ \citenamefont
  {Basov}}]{Sunku2021}%
  \BibitemOpen
  \bibfield  {author} {\bibinfo {author} {\bibfnamefont {S.~S.}\ \bibnamefont
  {Sunku}}, \bibinfo {author} {\bibfnamefont {D.}~\bibnamefont {Halbertal}},
  \bibinfo {author} {\bibfnamefont {T.}~\bibnamefont {Stauber}}, \bibinfo
  {author} {\bibfnamefont {S.}~\bibnamefont {Chen}}, \bibinfo {author}
  {\bibfnamefont {A.~S.}\ \bibnamefont {McLeod}}, \bibinfo {author}
  {\bibfnamefont {A.}~\bibnamefont {Rikhter}}, \bibinfo {author} {\bibfnamefont
  {M.~E.}\ \bibnamefont {Berkowitz}}, \bibinfo {author} {\bibfnamefont
  {C.~F.~B.}\ \bibnamefont {Lo}}, \bibinfo {author} {\bibfnamefont {D.~E.}\
  \bibnamefont {Gonzalez-Acevedo}}, \bibinfo {author} {\bibfnamefont {J.~C.}\
  \bibnamefont {Hone}}, \bibinfo {author} {\bibfnamefont {C.~R.}\ \bibnamefont
  {Dean}}, \bibinfo {author} {\bibfnamefont {M.~M.}\ \bibnamefont {Fogler}}, \
  and\ \bibinfo {author} {\bibfnamefont {D.~N.}\ \bibnamefont {Basov}},\ }\href
  {\doibase 10.1038/s41467-021-21792-2} {\bibfield  {journal} {\bibinfo
  {journal} {Nat. Comm.}\ }\textbf {\bibinfo {volume} {12}},\ \bibinfo {pages}
  {1641} (\bibinfo {year} {2021})}\BibitemShut {NoStop}%
\bibitem [{\citenamefont {Hubmann}\ \emph {et~al.}(2022)\citenamefont
  {Hubmann}, \citenamefont {Soul}, \citenamefont {Battista}, \citenamefont
  {Hild}, \citenamefont {Watanabe}, \citenamefont {Taniguchi}, \citenamefont
  {Efetov},\ and\ \citenamefont {Ganichev}}]{Hubmann2022}%
  \BibitemOpen
  \bibfield  {author} {\bibinfo {author} {\bibfnamefont {S.}~\bibnamefont
  {Hubmann}}, \bibinfo {author} {\bibfnamefont {P.}~\bibnamefont {Soul}},
  \bibinfo {author} {\bibfnamefont {G.~D.}\ \bibnamefont {Battista}}, \bibinfo
  {author} {\bibfnamefont {M.}~\bibnamefont {Hild}}, \bibinfo {author}
  {\bibfnamefont {K.}~\bibnamefont {Watanabe}}, \bibinfo {author}
  {\bibfnamefont {T.}~\bibnamefont {Taniguchi}}, \bibinfo {author}
  {\bibfnamefont {D.~K.}\ \bibnamefont {Efetov}}, \ and\ \bibinfo {author}
  {\bibfnamefont {S.~D.}\ \bibnamefont {Ganichev}},\ }\href {\doibase
  10.1103/physrevmaterials.6.024003} {\bibfield  {journal} {\bibinfo  {journal}
  {Phys. Rev. Materials}\ }\textbf {\bibinfo {volume} {6}},\ \bibinfo {pages}
  {024003} (\bibinfo {year} {2022})}\BibitemShut {NoStop}%
\bibitem [{\citenamefont {Deng}\ \emph {et~al.}(2020)\citenamefont {Deng},
  \citenamefont {Ma}, \citenamefont {Wang}, \citenamefont {Yuan}, \citenamefont
  {Watanabe}, \citenamefont {Taniguchi}, \citenamefont {Zhang},\ and\
  \citenamefont {Xia}}]{Deng2020}%
  \BibitemOpen
  \bibfield  {author} {\bibinfo {author} {\bibfnamefont {B.}~\bibnamefont
  {Deng}}, \bibinfo {author} {\bibfnamefont {C.}~\bibnamefont {Ma}}, \bibinfo
  {author} {\bibfnamefont {Q.}~\bibnamefont {Wang}}, \bibinfo {author}
  {\bibfnamefont {S.}~\bibnamefont {Yuan}}, \bibinfo {author} {\bibfnamefont
  {K.}~\bibnamefont {Watanabe}}, \bibinfo {author} {\bibfnamefont
  {T.}~\bibnamefont {Taniguchi}}, \bibinfo {author} {\bibfnamefont
  {F.}~\bibnamefont {Zhang}}, \ and\ \bibinfo {author} {\bibfnamefont
  {F.}~\bibnamefont {Xia}},\ }\href {\doibase 10.1038/s41566-020-0644-7}
  {\bibfield  {journal} {\bibinfo  {journal} {Nat. Photonics}\ }\textbf
  {\bibinfo {volume} {14}},\ \bibinfo {pages} {549} (\bibinfo {year}
  {2020})}\BibitemShut {NoStop}%
\bibitem [{\citenamefont {Kort-Kamp}\ \emph {et~al.}(2018)\citenamefont
  {Kort-Kamp}, \citenamefont {Culchac}, \citenamefont {Capaz},\ and\
  \citenamefont {Pinheiro}}]{KortKamp2018}%
  \BibitemOpen
  \bibfield  {author} {\bibinfo {author} {\bibfnamefont {W.~J.~M.}\
  \bibnamefont {Kort-Kamp}}, \bibinfo {author} {\bibfnamefont {F.~J.}\
  \bibnamefont {Culchac}}, \bibinfo {author} {\bibfnamefont {R.~B.}\
  \bibnamefont {Capaz}}, \ and\ \bibinfo {author} {\bibfnamefont {F.~A.}\
  \bibnamefont {Pinheiro}},\ }\href {\doibase 10.1103/physrevb.98.195431}
  {\bibfield  {journal} {\bibinfo  {journal} {Phys. Rev. B}\ }\textbf {\bibinfo
  {volume} {98}},\ \bibinfo {pages} {195431} (\bibinfo {year}
  {2018})}\BibitemShut {NoStop}%
\bibitem [{\citenamefont {Gao}\ \emph {et~al.}(2020)\citenamefont {Gao},
  \citenamefont {Li}, \citenamefont {Xin}, \citenamefont {Chen}, \citenamefont
  {Liu},\ and\ \citenamefont {Tian}}]{Gao2020a}%
  \BibitemOpen
  \bibfield  {author} {\bibinfo {author} {\bibfnamefont {X.-G.}\ \bibnamefont
  {Gao}}, \bibinfo {author} {\bibfnamefont {X.-K.}\ \bibnamefont {Li}},
  \bibinfo {author} {\bibfnamefont {W.}~\bibnamefont {Xin}}, \bibinfo {author}
  {\bibfnamefont {X.-D.}\ \bibnamefont {Chen}}, \bibinfo {author}
  {\bibfnamefont {Z.-B.}\ \bibnamefont {Liu}}, \ and\ \bibinfo {author}
  {\bibfnamefont {J.-G.}\ \bibnamefont {Tian}},\ }\href {\doibase
  10.1515/nanoph-2020-0024} {\bibfield  {journal} {\bibinfo  {journal}
  {Nanophotonics}\ }\textbf {\bibinfo {volume} {9}},\ \bibinfo {pages} {1717}
  (\bibinfo {year} {2020})}\BibitemShut {NoStop}%
\bibitem [{\citenamefont {Wang}\ \emph {et~al.}(2020)\citenamefont {Wang},
  \citenamefont {Bo}, \citenamefont {Ding}, \citenamefont {Wang},\ and\
  \citenamefont {Mu}}]{Wang2020}%
  \BibitemOpen
  \bibfield  {author} {\bibinfo {author} {\bibfnamefont {J.}~\bibnamefont
  {Wang}}, \bibinfo {author} {\bibfnamefont {W.}~\bibnamefont {Bo}}, \bibinfo
  {author} {\bibfnamefont {Y.}~\bibnamefont {Ding}}, \bibinfo {author}
  {\bibfnamefont {X.}~\bibnamefont {Wang}}, \ and\ \bibinfo {author}
  {\bibfnamefont {X.}~\bibnamefont {Mu}},\ }\href {\doibase
  10.1016/j.mtphys.2020.100238} {\bibfield  {journal} {\bibinfo  {journal}
  {Mater. Today Phys.}\ }\textbf {\bibinfo {volume} {14}},\ \bibinfo {pages}
  {100238} (\bibinfo {year} {2020})}\BibitemShut {NoStop}%
\bibitem [{\citenamefont {Chen}\ \emph {et~al.}(2022)\citenamefont {Chen},
  \citenamefont {Liu}, \citenamefont {Zeng},\ and\ \citenamefont
  {Li}}]{Chen2022}%
  \BibitemOpen
  \bibfield  {author} {\bibinfo {author} {\bibfnamefont {J.}~\bibnamefont
  {Chen}}, \bibinfo {author} {\bibfnamefont {C.}~\bibnamefont {Liu}}, \bibinfo
  {author} {\bibfnamefont {Z.}~\bibnamefont {Zeng}}, \ and\ \bibinfo {author}
  {\bibfnamefont {R.}~\bibnamefont {Li}},\ }\href {\doibase
  10.1103/physrevb.105.014309} {\bibfield  {journal} {\bibinfo  {journal}
  {Phys. Rev. B}\ }\textbf {\bibinfo {volume} {105}},\ \bibinfo {pages}
  {014309} (\bibinfo {year} {2022})}\BibitemShut {NoStop}%
\bibitem [{\citenamefont {Zheng}\ \emph {et~al.}(2022)\citenamefont {Zheng},
  \citenamefont {Song}, \citenamefont {Shan}, \citenamefont {Xin},\ and\
  \citenamefont {Cheng}}]{Zheng2022}%
  \BibitemOpen
  \bibfield  {author} {\bibinfo {author} {\bibfnamefont {Z.}~\bibnamefont
  {Zheng}}, \bibinfo {author} {\bibfnamefont {Y.}~\bibnamefont {Song}},
  \bibinfo {author} {\bibfnamefont {Y.~W.}\ \bibnamefont {Shan}}, \bibinfo
  {author} {\bibfnamefont {W.}~\bibnamefont {Xin}}, \ and\ \bibinfo {author}
  {\bibfnamefont {J.~L.}\ \bibnamefont {Cheng}},\ }\href {\doibase
  10.1103/physrevb.105.085407} {\bibfield  {journal} {\bibinfo  {journal}
  {Phys. Rev. B}\ }\textbf {\bibinfo {volume} {105}},\ \bibinfo {pages}
  {085407} (\bibinfo {year} {2022})}\BibitemShut {NoStop}%
\bibitem [{\citenamefont {Kim}\ \emph {et~al.}(2017)\citenamefont {Kim},
  \citenamefont {DaSilva}, \citenamefont {Huang}, \citenamefont {Fallahazad},
  \citenamefont {Larentis}, \citenamefont {Taniguchi}, \citenamefont
  {Watanabe}, \citenamefont {LeRoy}, \citenamefont {MacDonald},\ and\
  \citenamefont {Tutuc}}]{Kim2017a}%
  \BibitemOpen
  \bibfield  {author} {\bibinfo {author} {\bibfnamefont {K.}~\bibnamefont
  {Kim}}, \bibinfo {author} {\bibfnamefont {A.}~\bibnamefont {DaSilva}},
  \bibinfo {author} {\bibfnamefont {S.}~\bibnamefont {Huang}}, \bibinfo
  {author} {\bibfnamefont {B.}~\bibnamefont {Fallahazad}}, \bibinfo {author}
  {\bibfnamefont {S.}~\bibnamefont {Larentis}}, \bibinfo {author}
  {\bibfnamefont {T.}~\bibnamefont {Taniguchi}}, \bibinfo {author}
  {\bibfnamefont {K.}~\bibnamefont {Watanabe}}, \bibinfo {author}
  {\bibfnamefont {B.~J.}\ \bibnamefont {LeRoy}}, \bibinfo {author}
  {\bibfnamefont {A.~H.}\ \bibnamefont {MacDonald}}, \ and\ \bibinfo {author}
  {\bibfnamefont {E.}~\bibnamefont {Tutuc}},\ }\href {\doibase
  10.1073/pnas.1620140114} {\bibfield  {journal} {\bibinfo  {journal} {Proc.
  Natl. Acad. Sci.}\ }\textbf {\bibinfo {volume} {114}},\ \bibinfo {pages}
  {3364} (\bibinfo {year} {2017})}\BibitemShut {NoStop}%
\bibitem [{\citenamefont {Dantscher}\ \emph {et~al.}(2017)\citenamefont
  {Dantscher}, \citenamefont {Kozlov}, \citenamefont {Scherr}, \citenamefont
  {Gebert}, \citenamefont {B\"arenf\"anger}, \citenamefont {Durnev},
  \citenamefont {Tarasenko}, \citenamefont {Bel'kov}, \citenamefont
  {Mikhailov}, \citenamefont {Dvoretsky}, \citenamefont {Kvon}, \citenamefont
  {Ziegler}, \citenamefont {Weiss},\ and\ \citenamefont
  {Ganichev}}]{Dantscher2017}%
  \BibitemOpen
  \bibfield  {author} {\bibinfo {author} {\bibfnamefont {K.-M.}\ \bibnamefont
  {Dantscher}}, \bibinfo {author} {\bibfnamefont {D.~A.}\ \bibnamefont
  {Kozlov}}, \bibinfo {author} {\bibfnamefont {M.~T.}\ \bibnamefont {Scherr}},
  \bibinfo {author} {\bibfnamefont {S.}~\bibnamefont {Gebert}}, \bibinfo
  {author} {\bibfnamefont {J.}~\bibnamefont {B\"arenf\"anger}}, \bibinfo
  {author} {\bibfnamefont {M.~V.}\ \bibnamefont {Durnev}}, \bibinfo {author}
  {\bibfnamefont {S.~A.}\ \bibnamefont {Tarasenko}}, \bibinfo {author}
  {\bibfnamefont {V.~V.}\ \bibnamefont {Bel'kov}}, \bibinfo {author}
  {\bibfnamefont {N.~N.}\ \bibnamefont {Mikhailov}}, \bibinfo {author}
  {\bibfnamefont {S.~A.}\ \bibnamefont {Dvoretsky}}, \bibinfo {author}
  {\bibfnamefont {Z.~D.}\ \bibnamefont {Kvon}}, \bibinfo {author}
  {\bibfnamefont {J.}~\bibnamefont {Ziegler}}, \bibinfo {author} {\bibfnamefont
  {D.}~\bibnamefont {Weiss}}, \ and\ \bibinfo {author} {\bibfnamefont {S.~D.}\
  \bibnamefont {Ganichev}},\ }\href {\doibase 10.1103/PhysRevB.95.201103}
  {\bibfield  {journal} {\bibinfo  {journal} {Phys. Rev. B}\ }\textbf {\bibinfo
  {volume} {95}},\ \bibinfo {pages} {201103} (\bibinfo {year}
  {2017})}\BibitemShut {NoStop}%
\bibitem [{\citenamefont {Candussio}\ \emph {et~al.}(2021)\citenamefont
  {Candussio}, \citenamefont {Durnev}, \citenamefont {Slizovskiy},
  \citenamefont {J{\"o}tten}, \citenamefont {Keil}, \citenamefont {Bel'kov},
  \citenamefont {Yin}, \citenamefont {Yang}, \citenamefont {Son}, \citenamefont
  {Mishchenko}, \citenamefont {Fal'ko},\ and\ \citenamefont
  {Ganichev}}]{Candussio2021}%
  \BibitemOpen
  \bibfield  {author} {\bibinfo {author} {\bibfnamefont {S.}~\bibnamefont
  {Candussio}}, \bibinfo {author} {\bibfnamefont {M.~V.}\ \bibnamefont
  {Durnev}}, \bibinfo {author} {\bibfnamefont {S.}~\bibnamefont {Slizovskiy}},
  \bibinfo {author} {\bibfnamefont {T.}~\bibnamefont {J{\"o}tten}}, \bibinfo
  {author} {\bibfnamefont {J.}~\bibnamefont {Keil}}, \bibinfo {author}
  {\bibfnamefont {V.~V.}\ \bibnamefont {Bel'kov}}, \bibinfo {author}
  {\bibfnamefont {J.}~\bibnamefont {Yin}}, \bibinfo {author} {\bibfnamefont
  {Y.}~\bibnamefont {Yang}}, \bibinfo {author} {\bibfnamefont {S.-K.}\
  \bibnamefont {Son}}, \bibinfo {author} {\bibfnamefont {A.}~\bibnamefont
  {Mishchenko}}, \bibinfo {author} {\bibfnamefont {V.}~\bibnamefont {Fal'ko}},
  \ and\ \bibinfo {author} {\bibfnamefont {S.~D.}\ \bibnamefont {Ganichev}},\
  }\href {\doibase 10.1103/physrevb.103.125408} {\bibfield  {journal} {\bibinfo
   {journal} {Phys. Rev. B}\ }\textbf {\bibinfo {volume} {103}},\ \bibinfo
  {pages} {125408} (\bibinfo {year} {2021})}\BibitemShut {NoStop}%
\bibitem [{\citenamefont {Ziemann}\ \emph {et~al.}(2000)\citenamefont
  {Ziemann}, \citenamefont {Ganichev}, \citenamefont {Prettl}, \citenamefont
  {Yassievich},\ and\ \citenamefont {Perel}}]{Ziemann2000}%
  \BibitemOpen
  \bibfield  {author} {\bibinfo {author} {\bibfnamefont {E.}~\bibnamefont
  {Ziemann}}, \bibinfo {author} {\bibfnamefont {S.~D.}\ \bibnamefont
  {Ganichev}}, \bibinfo {author} {\bibfnamefont {W.}~\bibnamefont {Prettl}},
  \bibinfo {author} {\bibfnamefont {I.~N.}\ \bibnamefont {Yassievich}}, \ and\
  \bibinfo {author} {\bibfnamefont {V.~I.}\ \bibnamefont {Perel}},\ }\href
  {\doibase 10.1063/1.372423} {\bibfield  {journal} {\bibinfo  {journal} {J.
  Appl. Phys.}\ }\textbf {\bibinfo {volume} {87}},\ \bibinfo {pages} {3843}
  (\bibinfo {year} {2000})}\BibitemShut {NoStop}%
\bibitem [{\citenamefont {Drexler}\ \emph {et~al.}(2012)\citenamefont
  {Drexler}, \citenamefont {Dyakonova}, \citenamefont {Olbrich}, \citenamefont
  {Karch}, \citenamefont {Schafberger}, \citenamefont {Karpierz}, \citenamefont
  {Mityagin}, \citenamefont {Lifshits}, \citenamefont {Teppe}, \citenamefont
  {Klimenko}, \citenamefont {Meziani}, \citenamefont {Knap},\ and\
  \citenamefont {Ganichev}}]{Drexler2012}%
  \BibitemOpen
  \bibfield  {author} {\bibinfo {author} {\bibfnamefont {C.}~\bibnamefont
  {Drexler}}, \bibinfo {author} {\bibfnamefont {N.}~\bibnamefont {Dyakonova}},
  \bibinfo {author} {\bibfnamefont {P.}~\bibnamefont {Olbrich}}, \bibinfo
  {author} {\bibfnamefont {J.}~\bibnamefont {Karch}}, \bibinfo {author}
  {\bibfnamefont {M.}~\bibnamefont {Schafberger}}, \bibinfo {author}
  {\bibfnamefont {K.}~\bibnamefont {Karpierz}}, \bibinfo {author}
  {\bibfnamefont {Y.}~\bibnamefont {Mityagin}}, \bibinfo {author}
  {\bibfnamefont {M.~B.}\ \bibnamefont {Lifshits}}, \bibinfo {author}
  {\bibfnamefont {F.}~\bibnamefont {Teppe}}, \bibinfo {author} {\bibfnamefont
  {O.}~\bibnamefont {Klimenko}}, \bibinfo {author} {\bibfnamefont {Y.~M.}\
  \bibnamefont {Meziani}}, \bibinfo {author} {\bibfnamefont {W.}~\bibnamefont
  {Knap}}, \ and\ \bibinfo {author} {\bibfnamefont {S.~D.}\ \bibnamefont
  {Ganichev}},\ }\href {\doibase 10.1063/1.4729043} {\bibfield  {journal}
  {\bibinfo  {journal} {J. Appl. Phys.}\ }\textbf {\bibinfo {volume} {111}},\
  \bibinfo {pages} {124504} (\bibinfo {year} {2012})}\BibitemShut {NoStop}%
\bibitem [{Note1()}]{Note1}%
  \BibitemOpen
  \bibinfo {note} {Similar to our previous works \cite
  {Otteneder2020,Hubmann2022} (implementing much lower frequencies), in the
  absence of external bias we also observed polarization-dependent
  photocurrents. The corresponding results are presented in Sec.~\ref
  {appendix}}\BibitemShut {NoStop}%
\bibitem [{\citenamefont {Ganichev}\ and\ \citenamefont
  {Prettl}(2005)}]{Ganichev2005}%
  \BibitemOpen
  \bibfield  {author} {\bibinfo {author} {\bibfnamefont {S.~D.}\ \bibnamefont
  {Ganichev}}\ and\ \bibinfo {author} {\bibfnamefont {W.}~\bibnamefont
  {Prettl}},\ }\href {\doibase 10.1093/acprof:oso/9780198528302.001.0001}
  {\emph {\bibinfo {title} {Intense Terahertz Excitation of Semiconductors}}}\
  (\bibinfo  {publisher} {Oxford University Press},\ \bibinfo {address}
  {Oxford},\ \bibinfo {year} {2005})\BibitemShut {NoStop}%
\bibitem [{Note2()}]{Note2}%
  \BibitemOpen
  \bibinfo {note} {Note that the energy of photoexcited electrons and holes
  ($\sim $ \SI {120}{\milli \electronvolt }) is still insufficient for emission
  of optical phonons having higher energy of $\sim $ \SI {200}{\milli
  \electronvolt }.}\BibitemShut {Stop}%
\bibitem [{Note3()}]{Note3}%
  \BibitemOpen
  \bibinfo {note} {In general, the resistance $R$ depends on both electron and
  lattice temperatures, but at low $T$ the latter dependence is expected to be
  weak due to the lack of thermal phonons with relevant momenta.}\BibitemShut
  {Stop}%
\end{thebibliography}%

\end{document}